\documentclass[11pt]{article} % use larger type; default would be 10pt

\usepackage[utf8]{inputenc} % set input encoding (not needed with XeLaTeX)

\usepackage{geometry} % to change the page dimensions
\usepackage{hyperref}
\hypersetup{
    colorlinks=true,
    citecolor=cyan,
    linkcolor=.
}

\usepackage{graphicx} % support the \includegraphics command and options
\usepackage{xcolor}
\usepackage{parskip}
\usepackage{bm}
\usepackage{amsmath}
\usepackage{amssymb}
\usepackage{subfig}
\usepackage{footnpag}
\usepackage{caption}
\usepackage[symbol]{footmisc}

\makeatletter
\renewcommand*{\@biblabel}[1]{\hfill#1.}
\makeatother
\usepackage{natbib}
\setcitestyle{authoryear,citesep={; },notesep={; },round,aysep={,},yysep={;}}
\usepackage{indentfirst}
\usepackage{authblk}
\title{Europa's Hemispheric Color Dichotomy as a Constraint on Non-Synchronous Rotation}
\author[1]{Ethan R. Burnett\thanks{Corresponding author email: \texttt{ethan.burnett@colorado.edu}}}
\author[2,3]{Paul O. Hayne}
\affil[1]{Aerospace Engineering Sciences, University of Colorado Boulder}
\affil[2]{Astrophysical and Planetary Sciences, University of Colorado Boulder}
\affil[3]{Laboratory for Atmospheric and Space Physics, University of Colorado Boulder}

\date{} % Activate to display a given date or no date (if empty),
\begin{document}
\maketitle
\begin{abstract} Europa's surface reflectance exhibits a pronounced hemispheric dichotomy, which is hypothesized to form due to enhanced irradiation of the trailing hemisphere by energetic particles entrained in the jovian magnetosphere. We propose that this pattern can only persist if the timescale for discoloration is much shorter than that of Europa's rotation relative to the synchronous state, and provide a means for constraining the rotation rate using the observed color pattern. By decomposing the longitudinal ultraviolet and visible color variations from Voyager data into sine and cosine terms, we find no detectable signature of non-synchronous rotation (NSR). This same conclusion is reached with two observational models of discoloration: one representing an actively discoloring surface, and the other assuming that the present-day exogenic discoloration on the surface is in steady-state. Magnitudes of the expected signature are presented as functions of the age of the crater Pwyll, which is used to constrain the timescale of discoloration. Furthermore, we develop a physical model of discoloration to validate the geometric models, producing consistent results. The failure to identify a signature of NSR using Europa's hemispheric color dichotomy magnifies the outstanding problem of the origin of the stress to explain Europa's pervasive tectonic features.
\end{abstract}

\section{Introduction and Background}
The present-day icy surface of Europa is covered with large ridges and cracks, which are generally the result of tidal stresses during Europa's eccentric 3.55-day orbit about Jupiter. Tectonic models of surface cracks are only poorly constrained by observational data, primarily due to uncertainty in the thickness of Europa's ice shell, with most studies predicting thickness orders ranging from several kilometers up to tens of kilometers \citep{greenberg1998tectonic,nimmo2007global}. Some surface features are not well explained by the stress field predicted to arise from the diurnal variations in tidal forces alone. In particular, the works of \cite{HoppaScience1999} and \cite{Greenberg2003TidalStress,GreenbergEuropaRotation2002} suggest the existence of a large background stress field, which they attribute to slow non-synchronous rotation (NSR) from Europa's tidally locked orientation. Additionally, analysis by \cite{KATTENHORN2002490} and \cite{Geissler:1998Nature} also interpreted tectonic features as evidence for NSR. However, direct evidence of NSR is lacking, due to both the timescale of the hypothesized asynchronicity and the limitations of available data. Alternative models suggest that the background stress field could be due to ice shell thickening \citep{Nimmo2004} or polar wander \citep{Greenberg2003TidalStress}, which has support from several theoretical and observational studies \citep{SchenkEA:Nature2008, Matsuyama:2014, Sarid2002PolarWander,Ojakangas1989EuropaThermal,Ojakangas1989EuropaPW}.

Another salient feature of Europa's surface is the hemispheric color asymmetry, likely due to interactions between the surface and the jovian magnetosphere \citep{carlson2009europa}. Such irradiation patterns also appear to be common among the icy saturnian satellites \citep{Howett:2012PacMan}. Multispectral studies reveal a global asymmetry in albedo and color between the leading and trailing hemispheres, particularly in the relative ultraviolet (UV) spectral reflectance \citep{McEwen1986Europa}. Several studies have attributed the global hemispheric color asymmetry to an exogenic process \citep{MorrisonBurns1976, Nelson_Icarus1986}. The primary goal of this work is to investigate whether exogenic patterns in the color distribution of Europa's surface are consistent with NSR. 

\subsection{Exogenic Modification of Surface Ice}
Multiple studies have highlighted and studied the effects of the space environment on icy surfaces. Large-scale color patterns have been observed on the icy Galilean satellites and have been attributed to alteration by high-energy ions and electrons \citep{2009euro.book..529P}, as well as sulfur ion implantation \citep{Grundy:2007Science, Schenk:2011PlasmaSaturn}. 
The interaction between Jupiter's radiation environment and the surfaces of the Galilean satellites is discussed extensively in \cite{2004jpsm.book..485J}. In the Jupiter system, plasma co-rotates with the planet, overtaking the Galilean satellites in their orbital motion. As a result, plasma preferentially impacts the trailing hemispheres of these satellites. The dose rate at Europa's surface is hundreds of times the dose rate of the solar wind at the lunar surface, so this plasma environment is a significant factor in the surface chemistry of Europa. There are also impacts from high-energy particles and micrometeorites which mix the regolith, burying the products of radiolysis and bringing up fresh ice from below. Sputtering and sublimation act as additional resurfacing processes on the icy moons \citep{2009euro.book..529P}. A prominent feature resulting from the plasma environment is a bullseye pattern of higher UV absorption relative to violet at Europa, Ganymede, and Callisto \citep{2004jpsm.book..485J}.

The modification of icy surfaces by charged particles is a phenomenon that is not unique to the Galilean satellites. In particular, data from Cassini has provided a wealth of knowledge of similar processes affecting the saturnian moons. In \cite{Schenk:2011PlasmaSaturn}, global high-resolution color maps of Saturn's icy satellites Mimas, Enceladus, Tethys, Dione, and Rhea were generated from Cassini data. A decrease in albedo across the trailing hemispheres of Tethys, Dione, and Rhea was noted, which also corresponded with an increase in infrared (IR) to ultraviolet (UV) ratio (redness). The authors noted that the longitudinal variation of the IR/UV ratio is fit well by a Gaussian centered at the antapex of orbital motion, whereas a weaker enhancement in the IR/UV ratio was found in the leading hemispheres, but was fit well by a sine curve. They speculated that its origin was different from that of the trailing hemisphere enhancements. Overall, the hemispheric albedo and color patterns were attributed to the combined effects of magnetospheric plasma bombardment, E-ring particles, small heliocentric impactors, and sub-micron dust particles. They note that particles smaller than about $0.05 \ \mu\text{m}$ in radius behave like ions in Saturn's magnetic field, and circle the planet faster than orbital velocity, overtaking the moons from behind. Larger particles such as dust grains $0.5 \ \mu\text{m}$ in radius orbit 1.5\% faster than the Keplerian orbital velocity \citep{HAMILTON1993244}. 

A detailed example of exogenic modification of surface ice can be found in \cite{Howett:2012PacMan}. This work used data gathered from Tethys using Cassini's Composite Infrared Spectrometer (CIRS) and Imaging Subsystem (ISS). It identified a thermal and color anomaly with warmer nighttime and cooler daytime temperatures, and a darker appearance in the IR/UV color ratio maps. The temperature anomaly indicates an increase in thermal inertial in low latitudes of the leading hemisphere. Ice thermal inertia is proportional to $\sqrt{k\left(1 - p\right)}$, where $k$ is the thermal conductivity, and $p$ is porosity. The increase in thermal inertia inside the anomalous region indicates less porous and/or more thermally conductive surface ice than in the surrounding areas \citep{Howett:2012PacMan}. Because this anomaly is highly coincident with a region of high-energy electron bombardment, the authors concluded that the electrons alter the ice texture and make it more conductive than in the surrounding areas.

\section{Observations}
The hemispheric color asymmetry is a well-known surface feature of Europa. The color on the leading hemisphere (in the direction of the orbital velocity vector) is generally brighter than the trailing hemisphere, but the global asymmetry is most prominent in ultraviolet/violet spectral ratio measurements. Data from Voyager provide coincident UV and visible images over most of Europa's surface, whereas the UV data from Galileo were more limited spatially \citep{GREELEY1998}. This work therefore relies exclusively on Voyager data of the UV to violet albedo ratio. The prospects for further analysis using both Galileo and future datasets are discussed briefly in the Conclusions.

Work by \cite{Johnson_JGR1983} constructed multispectral mosaics of Europa from the Voyager images of Europa at UV ($0.35 \ \mu\text{m}$), violet ($0.41 \ \mu\text{m}$), blue ($0.48 \ \mu\text{m}$), and orange ($0.59 \ \mu\text{m}$) filters. The mosaics were constructed from a limited set of images at different phase angles and varying resolution (2 - 25 km/pixel), excluded the polar latitudes, but effectively covered the equatorial region and lower latitudes. Their work discusses various challenges encountered in constructing their mosaics to ensure consistent brightness and minimal disagreement between low and high-phase images. They compared spectra from their whole-disk images to telescopic spectra obtained previously \citep{McFadden1980} and noted general agreement except for a persistent reddening in the Voyager data, which was attributed to an error in the Voyager calibration. The UV/violet albedo ratio is less affected by phase angle than the absolute albedo, due to the similarity of photometric behavior at nearby wavelengths. Therefore, the mosaics of relative albedo are expected to be accurate despite differences in the phase angles of the measurements \citep{Johnson_JGR1983}.
Importantly, the authors noted that the trailing hemisphere has lower relative UV spectral reflectance than the leading hemisphere, for which plasma originating from Io's volcanism was previously suggested as the cause \citep{MorrisonBurns1976}. 

Work by \cite{Nelson_Icarus1986} took the Voyager global multispectral mosaic of Europa \citep{Johnson_JGR1983} and analyzed surface features with similar optical properties. In that work, the hemispheric asymmetry is highlighted in the ratio of ultraviolet to violet albedo, because the asymmetry becomes more pronounced with decreasing wavelength. The albedo ratio image also suppresses the effects of color variations from local surface features, revealing a smooth surface pattern of decreasing darkness as the cosine of the angle from the antapex of motion, or equivalently, a variation with the sine of the longitude in degrees East. While the variation in spectral properties of Europa's surface is likely affected by multiple processes both endogenic \citep{Trumbo_2019} and exogenic, the authors in \cite{Nelson_Icarus1986} argue convincingly that this particular simple sinusoidal pattern is consistent with an exogenic source, namely surface alteration by sulfur ion bombardment from the jovian magnetosphere. They note that as a consequence of their data processing techniques and the limited dataset, conclusions cannot be drawn from the high latitudes in the maps, so they generally focus on studying the albedo ratio in transects at lower latitudes. They also note slight mis-registration between the two component albedo images, which is important to consider when examining small-scale features. Similarly to  \cite{Johnson_JGR1983}, they discuss the reddening bias of the Voyager data in comparison to the telescopic data by \cite{McFadden1980}. 

In \cite{2004jpsm.book..485J}, the effects of high-energy electrons and sulfur ions on low-temperature ices were highlighted, and theoretical approximations of the longitudinal distribution of sulfur implantation on Europa's surface were provided. These results are characteristically similar to the sinusoidal variation seen in the ratio of ultraviolet to violet albedo in \cite{Nelson_Icarus1986}. This lends support to the hypothesis that a sinusoidal variation in discoloration could be caused by longitudinal variations in the intensity of plasma bombardment, which produces the observed discoloration patterns.

\begin{figure}[t!]
%\centering
\includegraphics[width=5.9in]{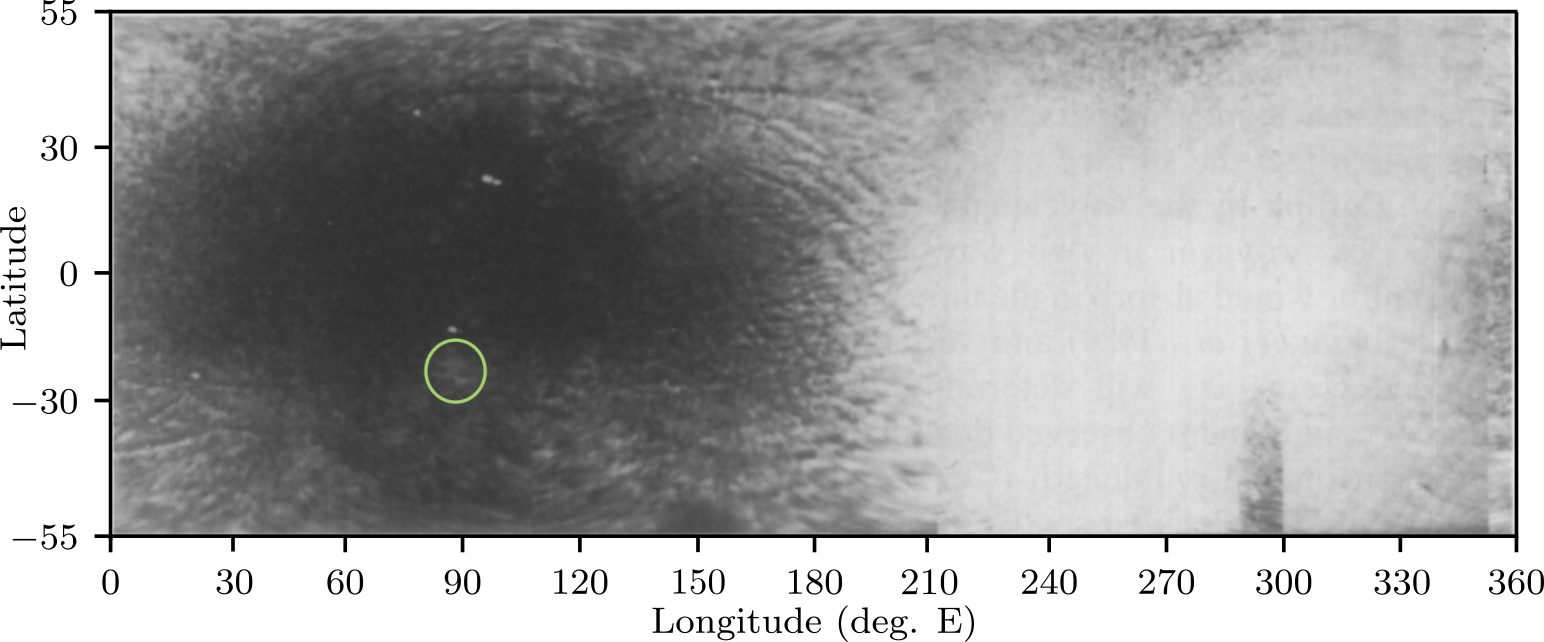} %UV2.png
\caption{Europa global UV/V albedo ratio map, obtained from Voyager data \citep{Nelson_Icarus1986}. The region of relatively diminished albedo ratio around Pwyll crater is circled.}
\label{fig:UV2}
\end{figure}
In \cite{Nelson_Icarus1986}, the albedo in violet and ultraviolet wavelengths was computed for fixed latitude for the whole range of longitude from the global multispectral mosaic and plotted in their Figure 14. The ``Ultraviolet/Violet" curve was obtained by the ratio of albedo in ultraviolet and violet, and is obtained from an average of $\pm 1^{\circ}$ latitude, and the ``Violet/Orange" curve is obtained similarly for the single row of pixels along the equator. The violet to orange albedo ratio was shown to reflect geologic variations. The Ultraviolet/Violet curve is reproduced digitally as the blue curve in Figure \ref{fig:FFit}. It shows a clear sinusoidal variation with longitude, which is also apparent in the global map of the ratio of ultraviolet to violet albedo, reproduced in Figure \ref{fig:UV2}. In Figure \ref{fig:UV2}, note that the faint streak traces remaining in the resultant image are caused by a slight misregistration between the two albedo images \citep{Nelson_Icarus1986}.

The presence of a clear maximum and minimum in the ultraviolet to violet albedo ratio transect require either an ongoing process not yet completed, or a process that has settled into equilibrium. \cite{Nelson_Icarus1986} argue that a completed process would produce a flatter top and bottom in the transect. Their hypothesis is that the global pattern is a result of the combined effects of sulfur ion implantation, micrometeorite gardening, and sputtering erosion -- all of which are due to exogenic sources.

%
%\newpage
%\clearpage

\section{Methods and Results}

Our approach is to model discoloration as a longitudinally varying process, and compare the predicted albedo patterns to those measured by Voyager. The exogenic flux of charged particles peaks on the trailing hemisphere, leading to a natural sinusoidal dependence of discoloration on longitude. Any NSR would act to smear this longitudinal pattern, and therefore, the observed pattern can in principle be used to infer the rate of NSR.

The \cite{Nelson_Icarus1986} data reproduced in Figure \ref{fig:FFit} is critical to this study. We seek to identify a signature of NSR in this sample longitudinal variation of the UV/V albedo ratio by first estimating the timescale of the discoloration process. This is done in two ways. First, because the UV/V ratio has not returned to background levels at Pwyll crater (see the green circled region in Figure \ref{fig:UV2}), we constrain the timescale of discoloration to be longer than the age of Pwyll crater. This is explored in Sections 3.1 and 3.2 with two similar simplified models of discoloration. The second method for estimating the timescale of discoloration is to use a physical model of the ice discoloration, provided in Section 3.3. These two approaches reach similar conclusions about the timescale of discoloration. 

Because our analysis relies heavily on the \cite{Nelson_Icarus1986} data in Figure \ref{fig:FFit}, it could be sensitive to systematic errors in the process for producing the original multispectral mosaics in \cite{Johnson_JGR1983}, or subsequent error introduced by the analysis in \cite{Nelson_Icarus1986}. Although photometric or data-processing effects could negatively impact the quality of the albedo data, no such effects are mentioned in \cite{Nelson_Icarus1986}, and indeed the authors commented on the highly sinusoidal nature of the UV/V albedo ratio. Because this analysis focuses only on the lowest-frequency signature in the data, it is expected to be fairly robust to localized errors. These issues are discussed further, along with other potential sources of error, in Section 4.

\subsection{Observational Models of Surface Discoloration}
With a longitudinally sinusoidal variation in the albedo ratio due to a sinusoidal variation in intensity of modification, for a slowly rotating body, a Fourier decomposition of the longitudinal variations in albedo should reveal a signature of the rotation history in the cosine component -- a remnant of discoloration in the past orientation. The following argument provides some analytical motivation for the existence of such a signature. 

\begin{figure}[h!]
\centering
\includegraphics[]{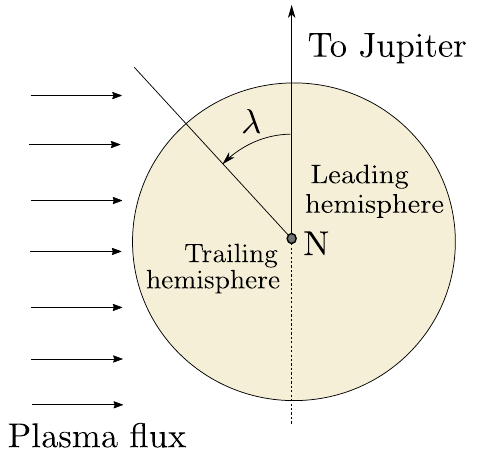}
\caption{Europa plasma bombardment, polar view. The longitude is measured in degrees East from the sub-jovian point. Plasma, slightly exceeding the local orbital velocity, preferentially impacts the trailing hemisphere of Europa, discoloring the icy surface. A slow rotation of the body from a tidally locked configuration will smear the resulting longitudinal pattern.}
\label{fig:UVpb}
\end{figure}

For a longitudinally varying optical parameter $r(\lambda)$ (such as the albedo ratio) arising from the plasma bombardment depicted in Figure \ref{fig:UVpb}, a simple dynamic model of alteration is proposed, given in Eq.~\eqref{rval5}.
 \begin{equation}
\label{rval5}
\dot{r}(\lambda,t) = f(\lambda) - n^{*}\frac{\partial r(\lambda, t)}{\partial \lambda}
\end{equation}
where $\lambda$ is the longitude in degrees east of the line towards Jupiter and $n^{*} = 2\pi/T^{*}$ is the angular rate of any NSR. This equation states that at a given time, the alteration rate is the sum of a term $f(\lambda)$ representing the longitudinally varying alteration rate of $r$, and a second term due to the assumption that the surface slowly rotates through a given angle $\lambda$. For tidally locked synchronous rotation, $n^{*} = 0$. Based on the analysis in \cite{Nelson_Icarus1986} and the nearly sinusoidal variation in sulfur ion flux predicted by \cite{2004jpsm.book..485J}, we specify $f(\lambda)$ to be sinusoidal:
 \begin{equation}
\label{rval6}
\dot{r}(\lambda,t) = a\left(\sin{\lambda} + 1\right) - n^{*}\frac{\partial r(\lambda, t)}{\partial \lambda}
\end{equation}
The term $a$ is one-half of $\dot{r}$ for a surface with normal vector $\hat{\bm{n}}$ satisfying $\hat{\bm{n}}\cdot\hat{\bm{u}} = -1$, where $\hat{\bm{u}}$ is the unit vector in the direction of Europa's velocity vector, or equivalently in the direction of the flow of the plasma.

The simple form of Eq. \eqref{rval6} is meant to serve as a first-order representation of a much more detailed process, which maximally affects the point on the equator facing the incoming plasma, and minimally affects the surface directly opposite on the equator, 180 degrees to the east. Any discoloration of the surface by ion bombardment is assumed to be attributed to two main effects. First, as more ions are embedded in the icy surface, the optical properties of the surface will be increasingly influenced by the presence of this material. Second, any post-bombardment small-scale structural or chemical changes to the ice could also alter the optical properties. For an ongoing discoloration process, it is thus assumed that the change in optical properties at some location will appear nearly linear over the time scale of multiple orbits around Jupiter, taking the time-averaged ion flux and resurfacing rate over many orbits. Therefore, $a$ is not considered to be a function of $t$ or $r$ in this analysis. 
\subsubsection{Model 1: Non-Saturating Discoloration Process}
One assumption of the form of Eq. \eqref{rval6} is that the discoloration process is non-saturating, and there does not exist a maximum value of $r$. One would expect the surface could be maximally altered by the plasma bombardment, or some steady-state discoloration could be achieved. However, an initial investigation with a non-saturating model is still instructive. 

Note that Europa's rotation is very slow, such that $n^{*} \ll 1$, and the differential equation is dominated by the external influence. Let $r(\lambda, t) = r^{0}(\lambda, t) + \delta r(\lambda, t)$ , where  $r^{0}(\lambda, t)$ is due to the external influence, and the rotation-induced deviation $\delta r(\lambda ,t)$ is small in comparison. Substituting this factorization into Eq. \eqref{rval6}:
\begin{equation}
\label{rval7}
\dot{r}^{0}(\lambda, t) + \delta\dot{r}(\lambda , t) = a\left(\sin{\lambda} + 1\right) - n^{*}\frac{\partial}{\partial \lambda}\left(r^{0}(\lambda, t) + \delta r(\lambda, t)\right)
\end{equation}
Subtracting $\dot{r}^{0}(\lambda,t) = a(\sin{\lambda} + 1)$ and noting that $\dfrac{\partial \delta r(\lambda, t)}{\partial \lambda} \ll \dfrac{\partial r^{0}(\lambda, t)}{\partial \lambda}$ for almost all longitudes except the critical points in $r^{0}(\lambda, t)$, the smallest term in Eq. \eqref{rval7} is dropped, and an approximate differential equation for the rotation-induced deviation is obtained: % except for the vicinity of $\lambda = 3\pi/2$
\begin{equation}
\label{rdev1}
\delta \dot{r} \approx -n^{*}\frac{\partial}{\partial \lambda}\left(r^{0}(\lambda, t)\right)
\end{equation}
where $r^{0}(\lambda, t) = r_{0}(\lambda) + a(\sin{\lambda} + 1)t$ and $r_{0}(\lambda)$ is an initial variation at $t = 0$. Substituting this into Eq. \eqref{rdev1} and applying the condition that the initial deviation is zero, we obtain $\delta r(\lambda, t)$, yielding the full solution for $r(\lambda, t)$:
\begin{equation}
\label{rtotal0}
r(\lambda, t) =  r_{0}(\lambda) + a(\sin{\lambda} + 1)t -\frac{1}{2}n^{*}t^{2}a\cos{\lambda} - \frac{\text{d}r_{0}(\lambda)}{\text{d}\lambda}n^{*}t
\end{equation}
Assuming no large variations in the initial uncontaminated distribution, $r_{0}(\lambda) \approx r_{0}$ for all longitudes $\lambda$, yielding:
\begin{equation}
\label{rtotal1}
r(\lambda, t) =  r_{0} + a(\sin{\lambda} + 1)t -\frac{1}{2}n^{*}t^{2}a\cos{\lambda}
\end{equation}

The result in Eq. \eqref{rtotal1} predicts that the NSR will produce a $\cos{\lambda}$ term in $r(\lambda, t)$. The sign of this term agrees with intuition. For an optical parameter that varies sinusoidally with longitude, the gradient is positive with longitude for $\lambda \in [-\pi/2, \pi/2]$, but negative for $\lambda~\in~[\pi/2, 3\pi/2]$. Thus, for a small counterclockwise rotation of the entire body, at a test longitude $\lambda'$ in the sub-jovian hemisphere, we would expect the local $r(\lambda', t)$ to be slightly below its predicted value for that longitude $\lambda'$ if the moon didn't rotate. Likewise, at an antipodal point in the opposite hemisphere, we would expect the local $r(\lambda'', t)$ to be slightly above its predicted value for that longitude $\lambda''$. Put more simply, if the discoloration rate increases with longitude from $\lambda = -\pi/2$ until $\lambda = \pi/2$ before decreasing with longitude from $\lambda = \pi/2$ until $\lambda = 3\pi/2$, then a small counterclockwise rotation will result in points in the sub-jovian hemisphere being continually a little bit less discolored than expected for a sinusoid, and the opposite hemisphere will be a little bit more discolored than expected. 
Eq. \eqref{rtotal1} agrees with this simple test.

\begin{figure}[h!]
\centering
\includegraphics[]{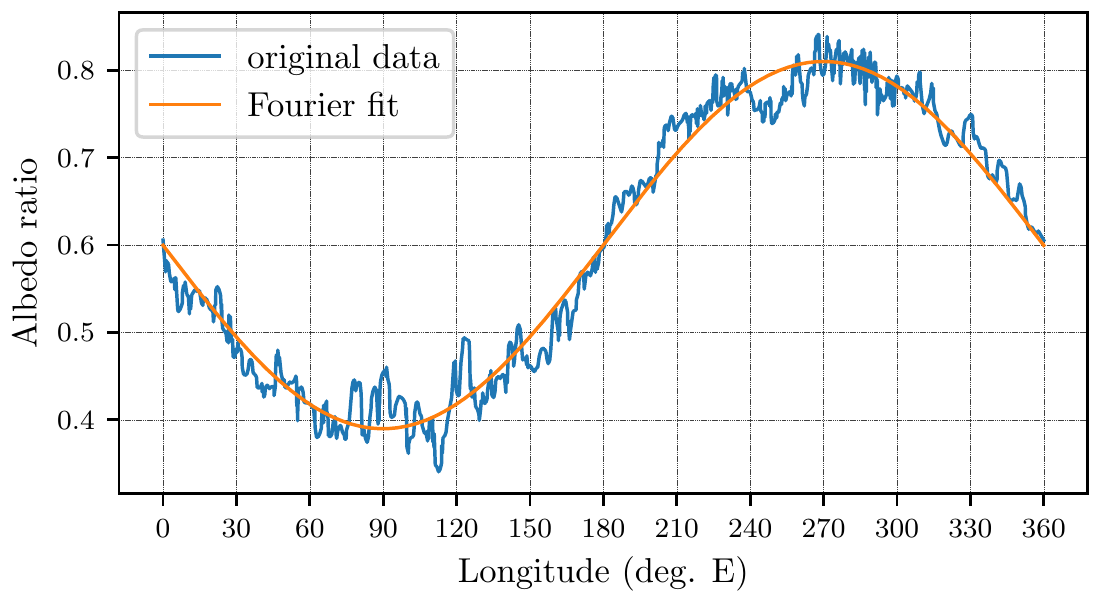} %albedoplot1.pdf
\caption{UV/V albedo ratio curve from \cite{Nelson_Icarus1986} based on an averaging of $\pm 1^{\circ}$ latitude transects of the global albedo ratio map constructed from Voyager data. A single-frequency Fourier fit of the data is superimposed.}
\label{fig:FFit}
\end{figure}

To use this first analytic model, Eq. \eqref{rtotal1} is equated to a Fourier series fit of the albedo ratio data from \cite{Nelson_Icarus1986} to determine the unknown parameter $n^{*}$ along with all other unknown parameters. The Fourier series fit is given in Eq. \eqref{r_FS1} where $a_{0}$, $a_{1}$, $b_{1}$ are the Fourier coefficients.
\begin{equation}
\label{r_FS1}
r = \frac{a_{0}}{2} + a_{1}\cos{\lambda} + b_{1}\sin{\lambda}
\end{equation}
Applying the Fourier decomposition to the data, the orange curve in Figure \ref{fig:FFit} is produced. The values of the Fourier coefficients are $a_{0} = 1.199586$, $a_{1} = -2.63\times 10^{-4}$, $b_{1} = -0.210128$. Note the very small ratio $a_{1}/b_{1}$. In the lowest frequency, the longitudinal distribution of the albedo ratio is almost entirely sinusoidal. 
\begin{figure}[h!]
\centering
\includegraphics[width=2.0in]{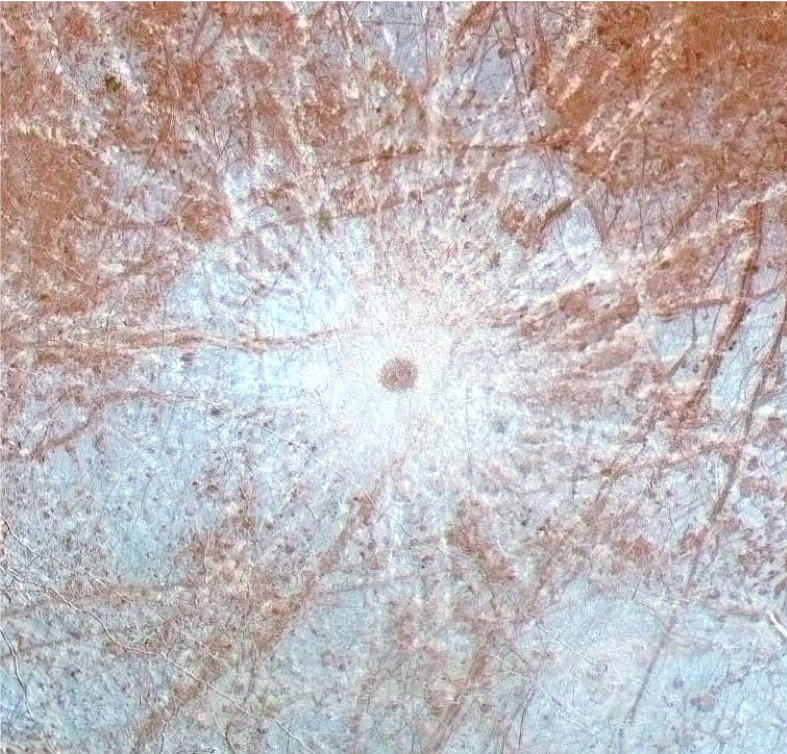}
\caption{Enhanced-color image of Pwyll crater, showing its bright rays superimposed on other surface features. This image was produced by combining low-resolution color data with a high-resolution mosaic of images from the Galileo spacecraft. The bright rays cross over many different types of terrain, and are thus inferred to be younger than anything they cross. From the Planetary Photojournal (Image ID: PIA01211)}
\label{fig:Pwyll}
\end{figure}

Applying the boundary conditions at $\lambda = 0, \ \pi/2, \ \pi, \ 3\pi/2$ to Eq. \eqref{rtotal1} and Eq. \eqref{r_FS1} and equating:
\begin{subequations}
\label{BCs1}
\begin{align}
r(\lambda = 0) = & \ r_{0} + a\left(t - \frac{1}{2}t^{2}n^{*}\right) = \frac{a_{0}}{2} + a_{1} \\
r(\lambda = \frac{\pi}{2}) = & \ r_{0} + 2at = \frac{a_{0}}{2} + b_{1}\\
r(\lambda = \pi) = & \ r_{0} + a\left(t + \frac{1}{2}t^{2}n^{*}\right) = \frac{a_{0}}{2} - a_{1} \\
r(\lambda = \frac{3\pi}{2}) = & \ r_{0} = \frac{a_{0}}{2} - b_{1}
\end{align}
\end{subequations}
Eq. \eqref{BCs1} yields the following underdetermined results relating the four parameters $r_{0}$, $a$, $n^{*}$, $t$:\footnote[2]{At this point, it is important to emphasize that the albedo ratio is a sinusoid, but it is flipped upside-down such that the point $\lambda = 90^{\circ}$ E in the trailing hemisphere corresponds to the lowest value of $r$. Thus, the maximal alteration in the analytic model actually corresponds to the lowest value in the data. This is not a problem, because instead of using Eq. \eqref{BCs1} to compute Eq. \eqref{BCs2}, the equation $r(\lambda) = r_{\text{max}} - r_{\text{meas}}(\lambda)$ can be used, where $r_{\text{max}}$ is the largest value in the albedo ratio curve, $r_{\text{meas}}(\lambda)$ is the Fourier fit of the albedo ratio data, and $r(\lambda)$ no longer incorporates the datum term $r_{0}$, by definition. Using this alternate approach induces no changes to the resulting equation for $n^{*}$. The only change is a sensible flip in the sign of $a$, such that $\eta = at = -b_{1}$. This is inconsequential to the analysis for computing $n^{*}$. The same property holds for all models in this paper: shiftings of the datum and flipping the curves for $r(\lambda)$ do not affect the analysis. However, the reader should note that any reference to ``maximal" alteration of the albedo ratio is a reference to the lowest values in Figure \ref{fig:FFit} and not the highest.}
\begin{subequations}
\label{BCs2}
\begin{align}
r_{0} = & \ \frac{a_{0}}{2} - b_{1} \\
\eta = & \ at = b_{1} \\
n^{*}t \sim & \ -2\frac{a_{1}}{b_{1}}
\end{align}
\end{subequations}

Estimating the timescale of formation $t_{P}$ of the observed color asymmetry can resolve the ambiguity in Eq. \eqref{BCs2} and enable the following order-of-magnitude estimate of $n^{*}$:
\begin{equation}
\label{n_star}
n^{*} \sim -2\frac{a_{1}}{t_{P}b_{1}}
\end{equation}
Pwyll crater, shown in Figure \ref{fig:Pwyll}, can provide some constraint on the scale of $t_{P}$, because the albedo ratio of the more recently exposed icy ejecta has not returned to surrounding background levels (see the circled region in Figure \ref{fig:UV2}).  Using an order-of-magnitude estimate of Pwyll's age \citep{BierhausIcarus2001} of $t_{P} \sim 10^{6}$ yr yields an approximate timescale of formation of the surface discoloration:
\begin{equation}
\label{T_star}
n^{*} \sim -2\frac{a_{1}}{t_{P}b_{1}} = -10^{-9} \ \text{rad}/\text{year} \ \rightarrow \ T^{*} \sim 10^{9} \ \text{years}
\end{equation}
Note that the negative sign of $n^{*}$ indicates a slightly sub-synchronous rotation rate. This is contradictory to the expected slightly super-synchronous result \citep{GoldreichPeale:1966Nature}, impyling that the NSR signal in the longitudinal distribution of discoloration is small enough that noise in the data completely obscures it, yielding the unexpected result $n^{*} < 0$.
If it can be shown that a noticeable signature in the discoloration data is expected for past suggested NSR rates, this result would imply that there is essentially no NSR and Europa is in fact tidally locked to Jupiter on geologically long timescales.

\subsubsection{Model 2: Exponentially Saturating Discoloration Process}
We developed a model of the discoloration dynamics in order to highlight the persistence of the important $n^{*}\cos{\lambda}$ term even for a saturating discoloration process, in which it is assumed that $\dot{r}$ decays exponentially by some unknown global alteration decay parameter $b$:
\begin{equation}
\label{m2eq1}
\dot{r}(\lambda, t) = \left(a\left(\sin{\lambda} + 1\right) - n^{*}\frac{\partial}{\partial \lambda}\left(r(\lambda, t)\right)\right)e^{-bt}
\end{equation}
We show in later sections that such a saturation process is predicted when the discoloration induced by contamination is in competition with erasive processes. Note that this equation supposes that the alteration rate decay constant $b$ does not vary significantly with longitude. Otherwise, the resulting global pattern would be distorted from the observed sinusoidal shape in the albedo ratio data. It is possible that the erasive processes can vary with longitude, which is explored in detail in Section 3.3.2.

Applying the same methods to solve Eq. \eqref{m2eq1} as were used with Eq. \eqref{rval7}, we obtain the rotation-invariant and largest component of the solution, $r^{0}(\lambda, t)~=~r_{0}(\lambda) + \frac{a}{b}(1 + \sin{\lambda})(1 - e^{-bt})$, then solve for the deviation term $\delta r(\lambda, t)$, and the complete approximate solution is obtained as:
\begin{equation}
\label{m2eqf}
\begin{split}
r(\lambda, t) = & \ r_{0}(\lambda) + \frac{a}{b}\left(1 + \sin{\lambda}\right)(1 - e^{-bt}) - n^{*}\bigg(\frac{a}{2b^{2}}\cos{\lambda} + \frac{1}{b}\frac{\text{d}r_{0}(\lambda)}{\text{d}\lambda} \\
& + \frac{1}{b^{2}}\left(\left(\frac{1}{2}ae^{-bt} - a\right)\cos{\lambda} - b\frac{\text{d}r_{0}(\lambda)}{\text{d}\lambda}\right)e^{-bt} \bigg)
\end{split}
\end{equation}
Applying the condition that the present-day distribution $r(\lambda)$ is expected to be in steady-state if it was generated by a saturating discoloration process, we let $e^{-bt} \rightarrow 0$, and the solution becomes:
\begin{equation}
\label{m2eqf2}
r(\lambda) = r_{0}(\lambda) + \frac{a}{b} + \frac{a}{b^{2}}\left(-\frac{1}{2}n^{*}\cos{\lambda} + b\sin{\lambda}\right) - \frac{n^{*}}{b}\frac{\text{d}r_{0}(\lambda)}{\text{d}\lambda}
\end{equation}
Similarly to Eqs. \eqref{BCs1} - \eqref{BCs2}, assuming $r_{0}(\lambda) = r_{0}$ and applying the boundary conditions to Eq. \eqref{m2eqf2} yields the following underdetermined results relating the four parameters $r_{0}$, $a$, $n^{*}$, $b$:
\begin{subequations}
\label{BCs2b}
\begin{align}
r_{0} = & \ \frac{a_{0}}{2}  - b_{1} \\
a = & b_{1}b \\
n^{*} = & \ -2\left(\frac{a_{1}}{b_{1}}\right)b
\end{align}
\end{subequations}
Pwyll crater can be used to provide an estimate for $b$. Using the longitude of Pwyll $\lambda_{P}$, the value $r_{P}$ from the albedo ratio map, and substituting an age for Pwyll $t_{P}$, the expression for $r(\lambda,t)$ can be used to solve for $b$, neglecting the small $n^{*}$ terms due to NSR:

\begin{equation}
\label{rPwyll}
r_{P} \approx r_{0} + b_{1}(1 + \sin{\lambda_{P}})(1 - e^{-bt_{P}})
\end{equation}

The signal due to NSR $\delta r(\lambda)$ is already expected to be quite small in comparison to $r^{0}(\lambda)$ due to the small value of $a_{1}/b_{1}$. Thus, to the precision to which $r_{P}$ is known, it is not necessary to account for the terms in $r(\lambda,t)$ that are linear in $n^{*}$. It can be easily shown that including those terms generally does not change the result appreciably.

Re-arranging Eq. \eqref{rPwyll} to solve for $b$ and substituting into Eq. \eqref{BCs2b}, the final equation for $n^{*}$ is:
\begin{equation}
\label{nstar_sat_final}
n^{*} \approx 2\frac{a_{1}}{b_{1}t_{P}} \text{ln}\left(1 - \frac{r_{p}-r_{0}}{b_{1}(1 + \sin{\lambda_{P}})}\right)
\end{equation}
Using $r_{0}$ from Eq. \eqref{BCs2b}, the previously obtained Fourier coefficients, and  $\lambda_{P} \approx 89^{\circ}$ E, this equation may be solved given $r_{P}$ and $t_{P}$. Comparing the albedo ratio of Pwyll to the known values of longitudes along the equator using the global map in \cite{Nelson_Icarus1986}, we obtain $r_{P}\approx 0.45$. Reusing the Pwyll crater age $t_{P} \sim 10^{-6}$, we obtain $n^{*} \sim -10^{-9}$, which agrees with the preceding analysis using the non-saturating model to within an order of magnitude. 

\begin{figure}[h!]
\centering
\includegraphics[]{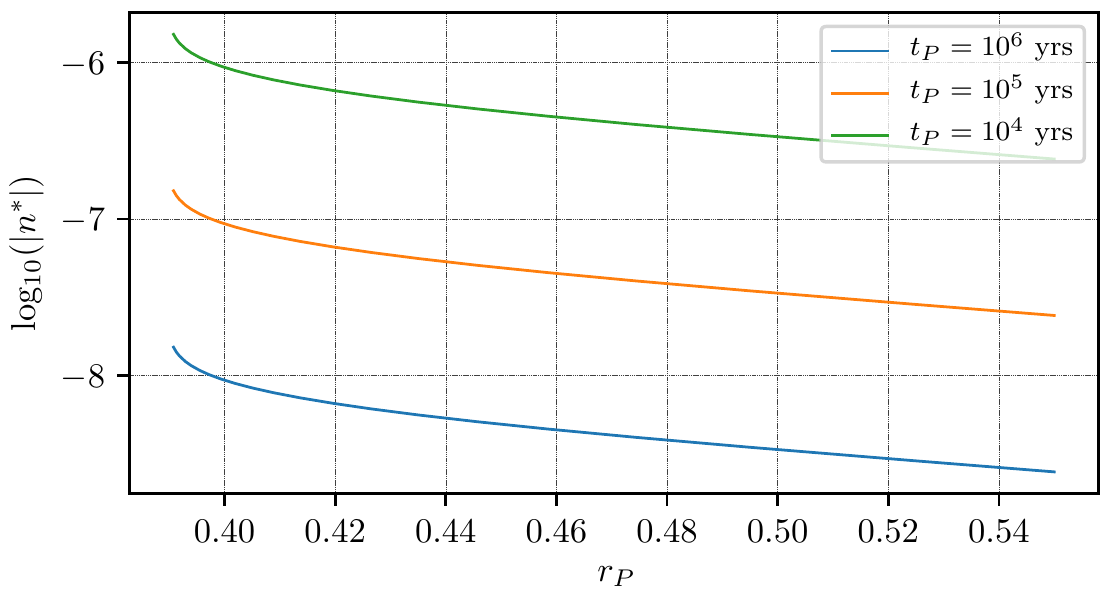}
\caption{Europa NSR rate, log\textsubscript{10}(yr\textsuperscript{$-1$}), vs. Pwyll region's age ($t_{P}$) and albedo ratio ($r_{P}$). These curves illustrate how the Pwyll-inferred NSR rate changes based on the assumed age of Pwyll and the value of its UV/V albedo ratio. Younger crater age implies faster NSR, and the albedo ratio of the Pwyll crater region exerts weak effect on the calculated NSR rate, unless it is near background saturation levels.}
\label{fig:nstar_from_rP}
\end{figure}

Note that there is disagreement in the age of Pwyll, and the value of $r_{P}$ can only be estimated from the figure in \cite{Nelson_Icarus1986}. It is important to investigate the consequences of both of these uncertainties. Using Eq. \eqref{nstar_sat_final}, a range of values for $r_{P}$, and test values of $t_{P}$, curves of $n^{*}$ vs. $r_{P}$ are obtained and given in Figure \ref{fig:nstar_from_rP}.

Changing the age of Pwyll does not yield $n^{*} > 0$; all values of $n^{*}$ referenced in Figure \ref{fig:nstar_from_rP} are negative. Furthermore, the curves of $n^{*}$ are relatively insensitive to the value of $r_{P}$ so long as the crater has not returned to the background saturation state (indicated by the lowest values of $r_{P}$). It is very clear from the albedo ratio map that the area in and around the crater differs from the surrounding region. Even if Pwyll is incredibly young ($10^{4}$ yrs), any value of $r_{P}$ greater than $0.4$ predicts a sub-synchronous timescale of $T^{*} > 10^{6}$ yrs. This analysis suggests that in order to yield a super-synchronous rotation period that is not on the scale of millions of years, the age of Pwyll must be very young, yielding a discoloration timescale that results in the NSR signature being small enough to be obscured by noise in the data in \cite{Nelson_Icarus1986}.

\subsection{Predicted Longitudinal Signatures for Given NSR Rates}

With the unexpected sub-synchronous rotation results, it is useful to ``work backwards", determining the expected values of $a_{1}/b_{1}$ for given values of $n^{*}$. As shown below, this analysis illustrates the detectable cosine signature in the longitudinal distribution to yield expected rates of NSR.

The non-saturating model provides the relationship
\begin{equation}
n^{*} \sim -2\frac{a_{1}}{t_{P}b_{1}}
\end{equation}
Re-arranging for $a_{1}$ as a function of $t_{P}$ and $n^{*}$:
\begin{equation}
\label{a1_bw_nonsat}
a_{1} \sim -\frac{1}{2}n^{*}t_{P}b_{1}
\end{equation}

The saturating model provides the relationship
\begin{equation}
n^{*} \sim -2\frac{a_{1}}{t_{P}b_{1}}\text{ln}\left(1 - \frac{r_{p}-r_{0}}{b_{1}(1 + \sin{\lambda_{P}})}\right)
\end{equation}
Re-arranging for $a_{1}$ as a function of $t_{P}$ and $n^{*}$, we obtain an expression similar to the non-saturating model, but with an additional term accounting for the particular value of $r_{P}$:
\begin{equation}
\label{a1_bw_sat}
a_{1} \sim -\frac{1}{2}n^{*}t_{P}b_{1}\cdot\left(\frac{-1}{\text{ln}\left(1 - \frac{r_{p}-r_{0}}{b_{1}(1 + \sin{\lambda_{P}})}\right)}\right)
\end{equation}

\begin{figure}[h!]
\centering
\includegraphics[]{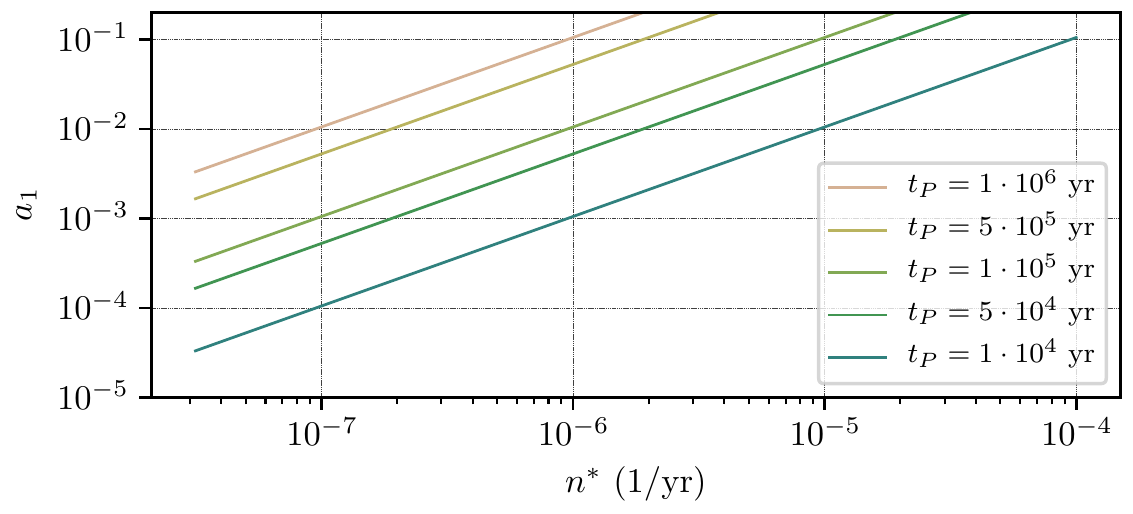}
\caption{Predicted cosine coefficient vs. NSR rate for the non-saturating model derived in Section 3.1.1. The cosine coefficient is the signature of NSR. Curves of different ages of Pwyll demonstrate that the older Pwyll is, the larger the expected cosine signature in the UV/V albedo ratio curve should be.}
\label{fig:nstar_backwards_nonsat}
\end{figure}

\begin{figure}[h!]
\centering
\includegraphics[]{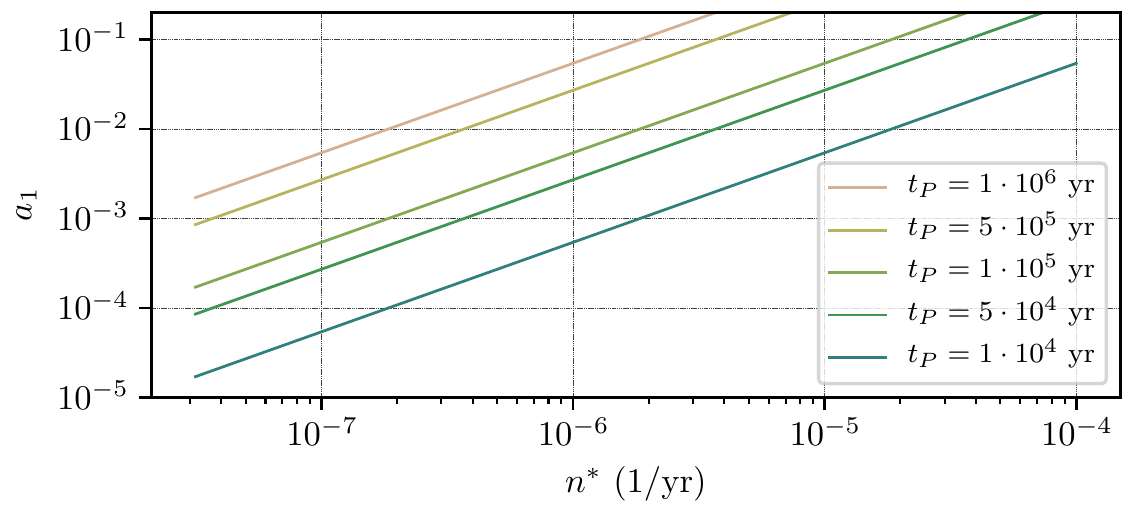}
\caption{Predicted cosine coefficient vs. NSR rate for the saturating model derived in Section 3.1.2. The cosine coefficient is the signature of NSR. Note all curves are shifted slightly down from their counterparts in Figure \ref{fig:nstar_backwards_nonsat}, as a saturating discoloration process will admit a more subtle NSR signature.}
\label{fig:nstar_backwards_sat}
\end{figure}

\begin{figure}[h!]
\centering
\includegraphics[]{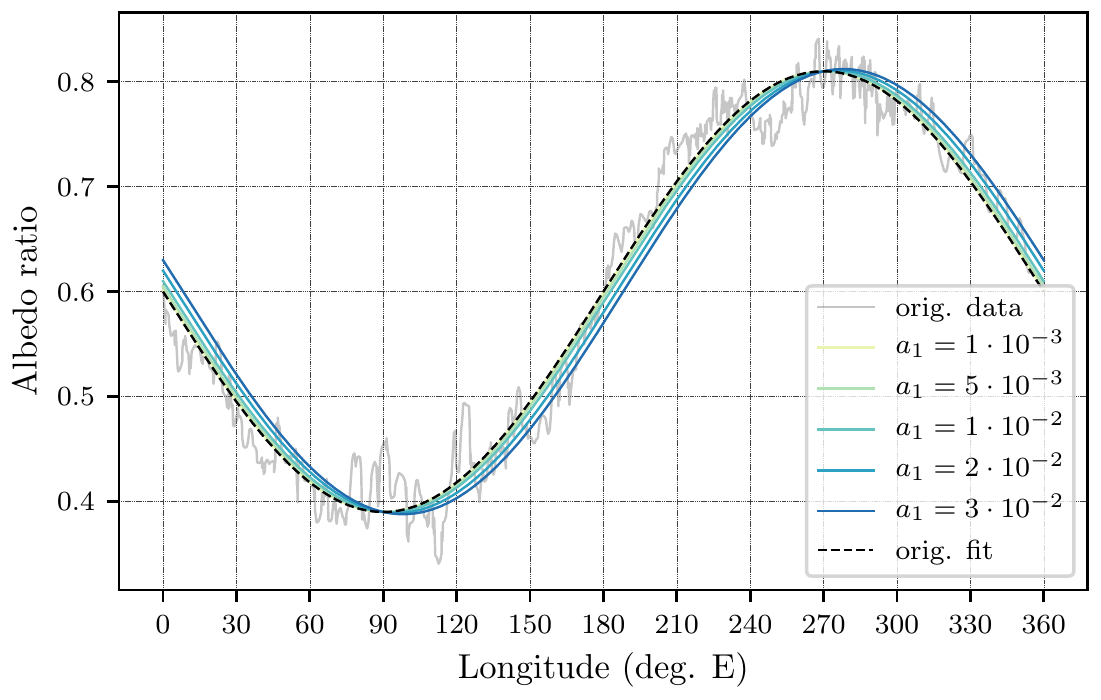}
\caption{Albedo ratio data and Fourier fits for selected values of $a_{1}$. This figure serves to illustrate various scales of NSR cosine signature in the Fourier fit of the original data.}
\label{fig:fourier_curve}
\end{figure}

Using the equations for $a_{1}$ predicted by the non-saturating and saturating discoloration models, it is possible to generate curves of the expected value of the NSR cosine signature as functions of $n^{*}$. The age of the crater Pwyll provides an additional free parameter, and we thus generate a family of curves. This is done for both the non-saturating model and the saturating model. The resulting coefficients are given as a function of $n^{*}$ in Figures \ref{fig:nstar_backwards_nonsat} and \ref{fig:nstar_backwards_sat}. As a reference, Figure \ref{fig:fourier_curve} shows the resulting change to the Fourier fit that would be achieved by various values of $a_{1}>0$.

The results generated by both models show that for values of $n^{*}$ corresponding to expected super-synchronous rotation rates, there should be a substantial signature in the longitudinal variation of albedo ratio, unless Pwyll is extraordinarily young. For example, for a $\sim 10^{5}$ yr period of NSR, these plots predict a considerable signal regardless of the age of Pwyll, for all ages tested $t_{P} > 10^{4}$ yrs. The Fourier coefficient $a_{1}$ should be opposite in sign and hundreds of times larger than the value obtained from the data in \cite{Nelson_Icarus1986}. This analysis suggests that either the crater Pwyll is extremely young ($10^{3}$ years or newer), or the NSR timescale is much slower than the period of $\sim 10^{5}$ yrs predicted by some studies \citep{Hoppa:Icarus2001, GreenbergEuropaRotation2002}.

%\newpage
%\clearpage
\subsection{Physical Model of Discoloration}
\subsubsection{Surface Ice Alteration}
The discoloration of Europa's surface is modeled as the result of a combination of processes acting on the ice at and near the surface \citep{2009euro.book..529P}. The appearance and spectral properties of an icy region are influenced by the concentration of non-ice contaminants, such as the sulfur ions that bombard the trailing hemisphere, and changes to the ice as a result of bombardment by high-energy particles \citep{carlson2009europa}. Meanwhile, impact gardening mixes fresh ice from further below, and these processes thus act in competition, with their relative strengths determining the properties of the ice. 

\begin{figure}[h!]
    \centering
    \includegraphics{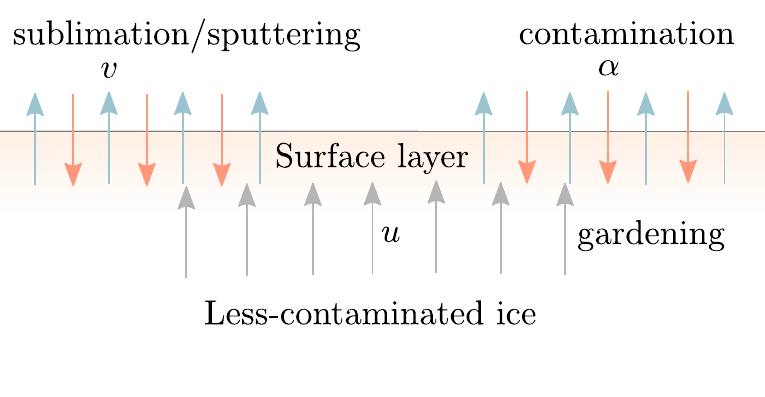} % v2
    \caption{A simple model of surface ice contamination. Impact gardening brings up fresh ice from below with overturn velocity $u$ at the lower boundary of a mixed surface layer. At the surface, sublimation and sputtering remove ice at a velocity $v$, and contamination occurs at a rate $\alpha$.}
    \label{fig:surface_ice_dyn}
\end{figure}

To approximate this phenomenon, consider a thin layer of gardening-mixed homogeneous surface ice of thickness $z$. On this layer, the combined effects of the three aforementioned processes are considered. See Figure \ref{fig:surface_ice_dyn}. First, impact gardening brings up fresh ice from below with overturn velocity $u$ at the lower boundary of the layer. At the surface, sublimation and sputtering remove ice at a velocity $v$, leaving behind non-ice contaminants. Let $M_{d}$ and $M_{i}$ denote the column mass of the contaminant and the ice (kg m\textsuperscript{$-2$}), and $q(t) = M_{d}/M_{i}$. Furthermore, $M_{d}/M_{i} \ll 1$, and we can use $M_{i} \approx \rho_{i}z$. The dynamics of $q$ are given by differentiation: 

\begin{equation}
\label{qdot1}
\begin{split}
\frac{\text{d}q}{\text{d}t} = & \ \frac{\text{d}}{\text{d}t}\frac{M_{d}}{M_{i}} = \frac{1}{M_{i}}\dot{M}_{d} - \frac{M_{d}}{M_{i}^{2}}\dot{M}_{i} \\
= & \ \frac{1}{\rho_{i}z}\left(\dot{M}_{d} - q\dot{M}_{i}\right)
\end{split}
\end{equation}
where $\dot{M}_{i} \approx -\rho_{i}v$ and the net effects of overturn on $M_{i}$ are assumed to be negligible. However, the gardening will noticeably dilute the contaminated ice in the surface layer.

Because the contaminant is only a small fraction of the composition of the ice layer, the contamination can be modeled as a process that happens to the ice in the layer, instead of an infiltrant flux of material, removing the need to introduce a distinct $\rho_{d}$ term. The full expression for $\dot{M}_{d}$ is a sum of the contamination process and the dilution by gardening:

\begin{equation}
    \label{Mddot}
    \dot{M}_{d} = \rho_{i}\alpha(q_{\text{max}} - q(t)) - \rho_{i}q(t)u
\end{equation}

where $q_{\text{max}}$ enforces that the contamination rate of the ice ceases as it achieves some maximal contamination, or as $q(t) \rightarrow q_{\text{max}}$, and $\alpha$ shares the units of $u$ and $v$. The form of this equation reflects the fact that the rate of contamination should be proportional to the available volume fraction that can still be contaminated. Substituting $\dot{M}_{i}$ and $\dot{M}_{d}$ into the equation for $\dot{q}$ and simplifying:

\begin{equation}
    \label{qdot2}
    \frac{\text{d}q}{\text{d}t} = \frac{1}{z}\left(\alpha\left(q_{\text{max}} - q(t)\right) - (u-v)q(t)\right)
\end{equation}

Integrating and applying the initial condition $q(0) = q_{0}$:

\begin{equation}
    \label{qoft1}
    q(t) = q_{0}e^{-\frac{\alpha + u - v}{z}t} + \frac{\alpha}{\alpha + u - v}q_{\text{max}}\left(1 - e^{-\frac{\alpha + u - v}{z}t}\right)
\end{equation}

As long as $\alpha + u - v \neq 0$, the solution is sensible and the resultant equilibrium is found by letting the transient terms decay to zero as $t \rightarrow \infty$:

\begin{equation}
    \label{qEndBehav}
    q \rightarrow \frac{\alpha}{\alpha + u - v}q_{\text{max}}
\end{equation}

Note that the equilibrium value of $q$ depends on the relative strengths of $\alpha$, $u$, and $v$.

\subsubsection{Equation for the Longitudinal Variation in Albedo Ratio}

We assume that the relationship between the concentration of contaminants in surface ice $q$ is linearly related to some optical parameter, $r = \xi q$. 
Additionally, allow the contaminant flux $\alpha$ to vary sinusoidally with longitude with maximum value $2\alpha_{1}$ and minimum value of zero. Likewise, let the sputtering and sublimation term $v$ be factored into a sinusoidally varying sputtering term of maximum value $2v_{1}$ and minimum value zero, and an isotropic sputtering and sublimation term $v_{2}$. Finally, to investigate the effect of anisotropy in impact gardening $u$, allow it to be a function of longitude as below:
\begin{equation}
\label{uMod1}
u(\lambda) = u(\lambda,w) = \overline{u}(-\sin{\lambda}+1)(1 - w) + \overline{u}w
\end{equation}
where $w \in [0, 1]$ is a parameter that determines the degree of symmetry in leading and trailing hemispheric impact gardening, and $\overline{u}$ scales the impact gardening rate function. If $w = 1$, $u = \overline{u}$ and the gardening is isotropic. As $w$ is decreased to zero, Eq. \eqref{uMod1} yields increasingly preferential gardening of the leading hemisphere, which is a possibility mentioned in works such as \cite{2004jpsm.book..485J}. 

Applying the aforementioned assumptions to the equation for $\dot{q}$ and adding the term due to rotation, the following differential equation is obtained:

\begin{equation}
    \label{rdotPDE1}
    \frac{\partial r(\lambda,t)}{\partial t} = c_{1}\bigg(\alpha_{\text{1}}(\sin{\lambda}+1)\big(r_{\text{max}}-r(\lambda,t)\big) - \Big(u(\lambda) - \big(v_{1}(\sin{\lambda} + 1) + v_{2}\big)\Big)r(\lambda,t)\bigg) - n^{*}\frac{\partial r(\lambda, t)}{\partial \lambda}
\end{equation}

where $c_{1} = 1/z$ and the linear correlation factor $\xi$ cancels from the equation.

Eq. \eqref{rdotPDE1} is a partial differential equation describing a behavior that is expected to saturate to a steady-state variation. Thus, as $t\rightarrow \infty$, $r(\lambda, t) \rightarrow r_{ss}(\lambda)$. Most importantly, $\dot{r} \rightarrow 0$, and as the discoloration dynamics saturate, and Eq. \eqref{rdotPDE1} reduces to an ordinary differential equation in longitude: 
\begin{equation}
    \label{rdotPDE2}
    c_{1}\bigg(\alpha_{\text{1}}(\sin{\lambda}+1)\big(r_{\text{max}}-r_{\text{ss}}(\lambda)\big) - \Big(u(\lambda) - \big(v_{1}(\sin{\lambda} + 1) + v_{2}\big)\Big)r_{\text{ss}}(\lambda)\bigg) = n^{*}\frac{\text{d} r_{\text{ss}}(\lambda)}{\text{d} \lambda}
\end{equation}

As before, note that the rotation is very slow, so $n^{*} \ll 1$. Also, let $r(\lambda, t) = r^{0}(\lambda,t) + \delta r(\lambda, t)$, where $r^{0}(\lambda,t)$ is due exclusively to the external influence, and the rotation-induced deviation $\delta r(\lambda, t)$ is very small in comparison. Applying this assumption to the steady-state distribution:

\begin{equation}
    \label{rSSnew1}
    r_{ss}(\lambda) = r_{ss}^{0}(\lambda) + \delta r_{ss}(\lambda)
\end{equation}

The term $r^{0}_{ss}(\lambda)$ is found by solving Eq. \eqref{rdotPDE2} algebraically when $n^{*} = 0$, obtaining the following:\footnote[2]{Because this derivation could be exactly repeated using the alternate definition $q = (r - r_{0})/\xi$, the general solution can also have an arbitrary constant term $r_{0}$ added.}

\begin{equation}
    \label{rss0a}
    r_{ss}^{0}(\lambda) = r_{\text{max}}\left(\sin{\lambda} + 1\right)\cdot\frac{\alpha_{1}}{(2-w)\overline{u} - v_{2} + (\sin{\lambda} + 1)(\alpha_{1} - v_{1} - \overline{u}(1-w))}
\end{equation}

Because the present-day (steady-state) variation in $r(\lambda)$ is very nearly sinusoidal, the following constraint on the relative process rates can be determined from above:
\begin{equation}
    \label{Constr1}
    |(2-w)\overline{u} - v_{2}| \gg |\alpha_{1} - v_{1} - (1-w)\overline{u}|
\end{equation}
Because the impact gardening rate $u$ dominates the other process rates \citep{2009euro.book..529P}, Eq. \eqref{Constr1} implies that either the impact gardening and surface contamination are in a state of balance, $\alpha_{1} \approx (1 - w)\overline{u}$, or that $w \approx 1$, indicating a mostly isotropic impact gardening influence. Either of these conditions would yield the resultant albedo ratio distribution observed at the present day.

Recall that $u$ is impact gardening, $v_{2}$ is sublimation and isotropic sputtering, $\alpha_{1}$ is due to sulfur ion implantation, and $v_{1}$ is associated with the sinusoidally varying component of the sputtering.

\begin{figure}[h!]
    \centering
    \includegraphics{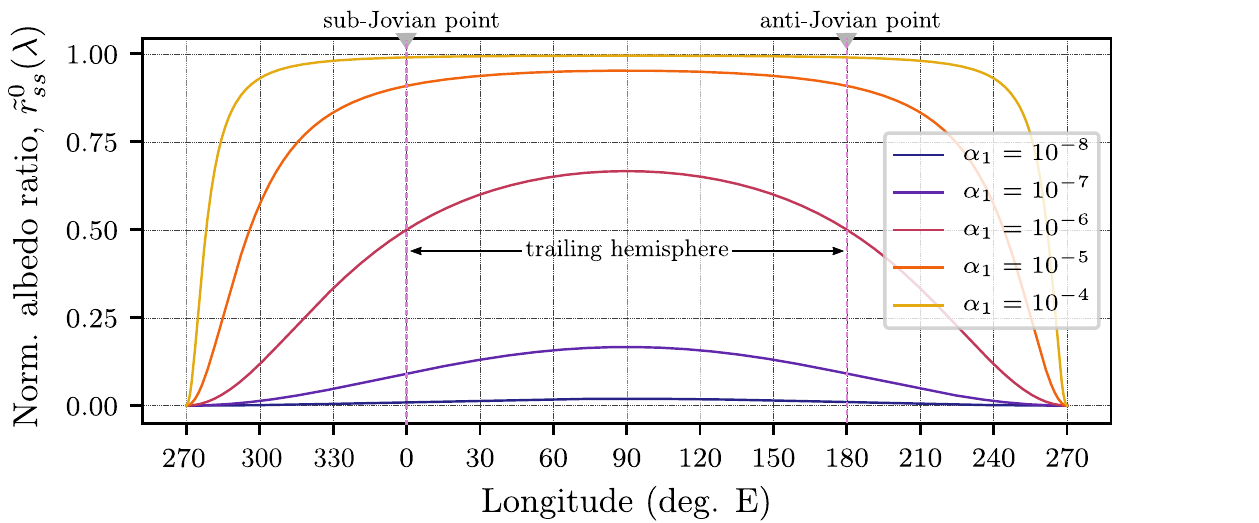}
    \caption{Global steady-state patterns in albedo ratio, normalized. Fixing the rates of impact gardening and sublimation/sputtering, the contaminant flux rate $\alpha$ is varied from $10^{-8}$ to $10^{-4}$ m/yr, producing a family of curves between minimal and maximal alteration. The curves for lower values of $\alpha_{1}$ are similar to the observed sinusoidal UV/V albedo ratio variation in the Voyager data.}
    \label{fig:global_curves}
\end{figure}

It is useful to explore the behavior of the steady-state rotation-invariant component of the solution $r^{0}_{ss}(\lambda)$ as the relative strength of $\alpha_{1}$ and $u$ varies. To do this, we consider the variation in the normalized curve $\tilde{r}^{0}_{\text{ss}}(\lambda)$ for which $r_{\text{max}} \equiv 1$. Let $w = 1$, and \cite{2009euro.book..529P} provide values for $u = w\overline{u}$ and $v_{2}$. We fix the values $u = 10^{-6}$ m/yr, $v_{1} = 8\times 10^{-9}$ m/yr (assumed), and $v_{2} \sim 1.6 \times 10^{-8}$ m/yr, and vary the value of $\alpha_{1}$, producing the plot in Figure \ref{fig:global_curves}.

Figure \ref{fig:global_curves} shows that as the ratio $\alpha_{1}/u$ increases and the relative strength of plasma bombardment to impact gardening increases, the trailing hemisphere (centered) becomes increasingly altered. It also increasingly approaches a flat distribution, and the transition to the unaffected point in the center of the leading hemisphere becomes more sharp. The flattening of the global distribution as the contamination rate dominates the gardening rate ($\alpha_{1}/u_{1} \gg 1$) matches the behavior predicted in \cite{Nelson_Icarus1986}. However, the observed albedo ratio pattern is sinusoidal in longitude, so the surface ice does not approach this theoretical high degree of alteration.

To solve for the small variation $\delta r_{ss}(\lambda)$, the steady-state terms are substituted into Eq. \eqref{rdotPDE2}, and the term $n^{*}\frac{\text{d}\delta r_{ss}(\lambda)}{\text{d}\lambda}$ is neglected as in prior derivations due to the dominance of $n^{*}\frac{\text{d} r^{0}_{ss}(\lambda)}{\text{d}\lambda}$. The following approximate result is obtained:

\begin{equation}
    \label{drssa}
    \delta r_{ss}(\lambda) \approx - n^{*}\frac{\text{d}}{\text{d}\lambda}\left(r^{0}_{ss}(\lambda)\right)\cdot\frac{z}{(2-w)\overline{u} - v_{2} + (\sin{\lambda} + 1)(\alpha_{1} - v_{1} - \overline{u}(1-w))}
\end{equation}

Because $r^{0}_{ss}(\lambda)$ is approximately sinusoidal, this result predicts that NSR will result in a cosine signature in the longitudinal variation of albedo ratio. Furthermore, the sign of this term agrees with the  lower-fidelity non-saturating and saturating models.

Substituting the Fourier series fit of $r_{ss}(\lambda) = a_{0}/2 + a_{1}\cos{\lambda} + b_{1}\sin{\lambda}$ into the above equation for the deviation, and neglecting higher-order and sub-dominant terms, the following approximate equation is obtained:

\begin{equation}
    \label{n_star_new1}
    a_{1} \approx -n^{*}b_{1}\frac{z}{(2 - w)\overline{u} -v_{2}}
\end{equation}

Re-arranging for $n^{*}$:

\begin{equation}
    \label{n_star_new2} 
    n^{*} \approx -\frac{a_{1}}{b_{1}}\left(\frac{(2 - w)\overline{u}-v_{2}}{z}\right)
\end{equation}

Thus, $n^{*}$ can be estimated in terms of the known Fourier coefficients $a_{1}$ and $b_{1}$, the gardening rate $u$, the sublimation rate $v_{2}$, the surface ice layer thickness $z$, and the undetermined weight $w \in [0, 1]$. For an order-of magnitude analysis, the $(2 - w)$ term is unimportant, and we can treat it as unity. Note that this approximate expression is conveniently not a function of the sulfur ion implantation rate parameter $\alpha_{1}$, because both $r^{0}_{ss}(\lambda)$ and $\delta r(\lambda)$ are approximately linear in $\alpha_{1}$. We reuse the values $u \sim 10^{-6}$ m/yr and $v_{2} \sim 1.6 \times 10^{-8}$ m/yr, along with the values of the Fourier coefficients $a_{1}$ and $b_{1}$ from Section 3.1.1. This yields $n^{*} \sim 10^{-9}/z$, where $z$ is the depth of the thin homogeneous surface layer. Note that $z$ is zero for freshly extruded ice and becomes deeper as the ice is older. This happens on geologic timescales \citep{Costello:LPSC2020, COSTELLO2018327}.

While the observational models use Pwyll to effectively determine the timescale of discoloration and constrain the value of $n^{*}$, the physical model provides $n^{*}$ as a function of the ratio between the gardening and isotropic sputtering/sublimation rate and the depth of the surface layer $z$, independent of Pwyll's age. Determining the depth of the homogeneous surface layer thus becomes vitally important for making quantitative conclusions using the physical model. 

The mean stopping depth for energetic electrons is $\sim$1 mm \citep{2004jpsm.book..485J}, while sulfur ions are stopped even sooner. Impact gardening affects the surface to greater depths, driving contaminants deeper and thoroughly mixing the surface layer. It has been predicted that at least the top 10 cm of Europa's ice shell is thoroughly mixed, suggesting that the scale of the homogeneous surface layer should be on the order of centimeters \citep{Costello:LPSC2020, COSTELLO2018327}.

\subsubsection{Predicted Longitudinal Signatures}
\begin{figure}[h!]
\centering
\includegraphics[]{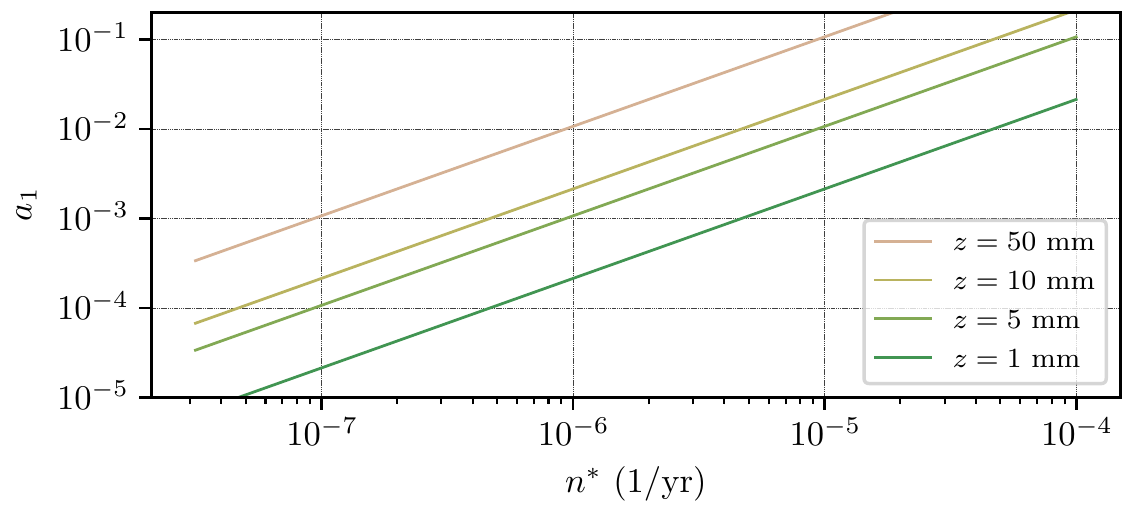}
\caption{Cosine coefficient vs. NSR rate for the contamination-based model derived in Section 3.3.2. The cosine coefficient is the signature of NSR. This is a similar result to Figures \ref{fig:nstar_backwards_nonsat} and \ref{fig:nstar_backwards_sat}, except that the depth of the homogenous mixing layer is the uncertain parameter, instead of the age of Pwyll crater.}
\label{fig:nstar_backwards_phys}
\end{figure}

Here the analysis from Section 2.3 is repeated with the new physical model. The cosine term in the Fourier fit can be expressed as a function of $n^{*}$, the depth of the control layer $z$, and other physical parameters:

\begin{equation}
a_{1} \approx -n^{*}b_{1}\frac{z}{u-v_{2}}
\end{equation}

where an order-of-magnitude analysis permits writing $u$ instead of $(2 - w)\overline{u}$. Using the above equation for $a_{1}$, curves of the expected value of the NSR cosine signature are generated as functions of $n^{*}$ in Figure \ref{fig:nstar_backwards_phys}. The depth of the surface ice layer $z$ is a free parameter. Recall that reasonable values are expected to be on the order of $10^{-2}$ m, corresponding to the top two curves in the plot. Note that for NSR on the timescale of 0.1 Myr ($n^{*} \sim 10^{-5}$), these results predict that the parameter $a_{1}$ should be at least $\sim 100$ times larger than the absolute value of $10^{-4}$ obtained in the Fourier analysis of the albedo ratio data, and should also be positive instead of negative. Even for a NSR timescale of 1 Myr, we expect $a_{1}$ to be on the order of $10^{-3}$ to $10^{-2}$.

\subsubsection{Similarities Between Physical and Saturation Models}
Recall the form of $q(t)$ for a small region is given by Eq.~\eqref{qoft1}.
Differentiating this equation, the resulting equation may be expressed in the following form:
\begin{equation}
    \label{rdot_derivgen1}
    \dot{q} = de^{-ft}
\end{equation}
where $d$ is a bulk alteration rate constant incorporating $\alpha$, $u$, and $v$, and $f = \frac{\alpha + u - v}{z}$ is a decay rate. Because it has already been determined that the impact gardening rate $u$ dominates the longitudinally varying rates $\alpha$ and $v$, we can say $f \sim u/z$, which will not vary significantly with longitude if $u$ is isotropic. We have explicitly assumed $q$ and $r$ are linearly correlated and that the rotation-independent component of the saturating model was of the form $\dot{r} = a\left(\sin{\lambda} + 1\right)e^{-bt}$, where the alteration rate varies with longitude, but the decay rate does not. The appearance of the decaying exponential in the saturating model thus corresponds to the physical model in the case that impact gardening is isotropic. Note that the true impact gardening effect is expected to have some longitudinal variation, as discussed in Section 3.3.2.

\section{Discussion}
\subsection{Summary of Results}

For this study, we test the behavior of three models of exogenic discoloration on Europa and resulting longitudinal variations in albedo ratio near the equator. All three models failed to find a signature of non-synchronous rotation (NSR) because the Fourier coefficients computed satisfy $a_{1}/b_{1} > 0$, which would correspond to an extremely slow and nonphysical sub-synchronous rotation $n^{*} < 0$ using Eqs.~\eqref{n_star}, \eqref{nstar_sat_final}, and \eqref{n_star_new2}. Then, functional curves of $a_{1}$ vs. supersynchronous values of $n^{*}$ are plotted for select values of either the age of Pwyll $t_{P}$ or the depth of the homogeneous surface ice layer $z$. These results are given in Figures \ref{fig:nstar_backwards_nonsat},  \ref{fig:nstar_backwards_sat}, and  \ref{fig:nstar_backwards_phys}. The results in Figures \ref{fig:nstar_backwards_nonsat} and  \ref{fig:nstar_backwards_sat} show that for values of $n^{*}> \mathcal{O}(10^{-7})$ yr\textsuperscript{$-1$}, there should be a substantial signature in the longitudinal variation of albedo ratio, unless Pwyll is very young (well under a million years old). Similarly, the results in Figure \ref{fig:nstar_backwards_phys} predict a substantial NSR signature for values of $n^{*}> \mathcal{O}(10^{-6})$ yr\textsuperscript{$-1$} or so, unless the homogeneous surface ice layer is unexpectedly thin. The results agree to within an order of magnitude and seem to especially reject the possibility of $n^{*}> \mathcal{O}(10^{-5})$ yr\textsuperscript{$-1$}. Here, the sources of uncertainty and error in these results are discussed and explored in depth to try and establish the likelihood of various constraints on the magnitude of $n^{*}$.

\subsection{The Age of Pwyll Crater}
Pwyll crater is the youngest crater of the 10-km scale on Europa. A crater of the size of Pwyll is expected to form on Europa approximately every 4 Myr \citep{Levison:Icarus2000}. Here, we use a probabilistic approach to gain more information about the likely age of Pwyll. 

\begin{figure}[h!]
\centering
\includegraphics[]{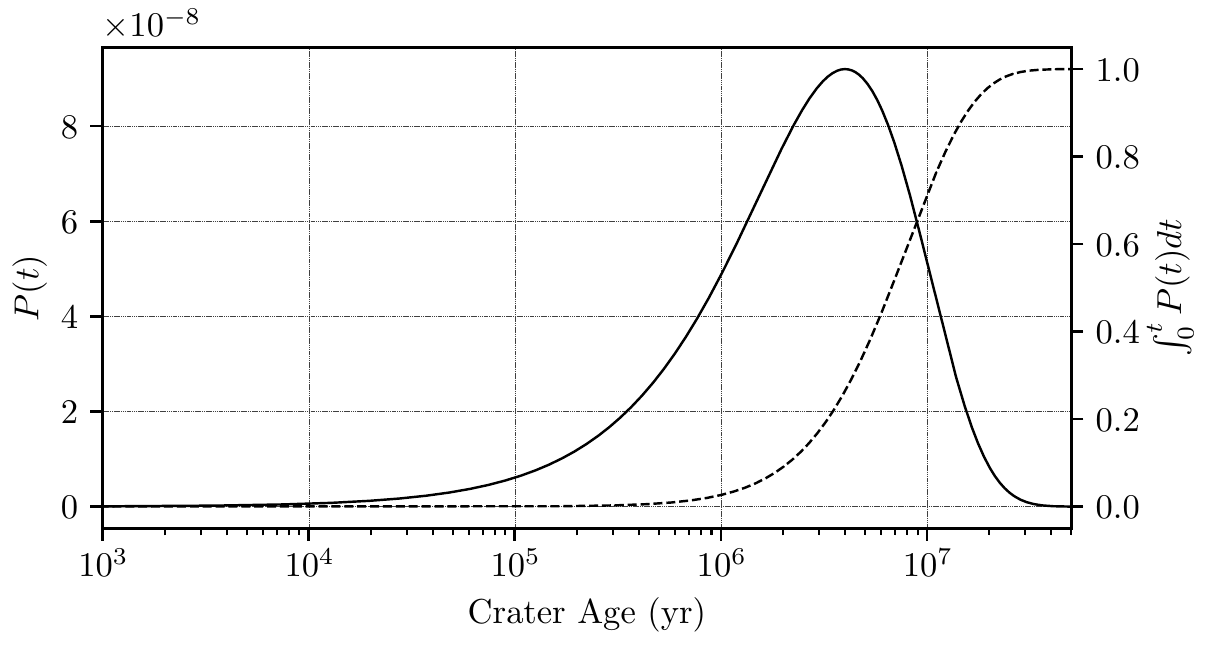}
\caption{Probability distribution of Pwyll-sized crater age. Provided are the probability distribution function (solid line) and cumulative distribution function (dashed line) for the age of a Pwyll-sized crater on the surface, assuming a Poisson process with $4\times 10^{6}$ yrs on average between impacts.}
\label{fig:PwyllProb}
\end{figure}

The Poisson distribution can be modified to determine the probability of $k$ discrete events in an interval $t$ for a memoryless process with average event rate $\gamma = 1/\tau$ where $\tau$ is the average time between events:
\begin{equation}
\label{Poisson1}
P(k \ \text{events in interval} \ t) = \frac{\left(\gamma t\right)^{k}e^{-\gamma t}}{k!}
\end{equation}
Eq. \eqref{Poisson1} can be used to determine the probability in an interval of time that a certain number of impacts will occur, given a known impact frequency. However, examining this expression as a function of time $t \in [0, \infty]$ will yield the relative probabilities of $k$ events occurring during specified intervals in that range. Setting $k = 1$, a probability density function $P(t)$ for single crater formation is derived via normalization of the relative probability expression $P_{u}$:
\begin{equation}
\label{Poisson2}
P(t) = lP_{u}(t) = l\gamma te^{-\gamma t}
\end{equation}
\begin{equation}
\label{Poisson3}
\int_{0}^{\infty} P_{u}(t) \text{d}t = \int_{0}^{\infty}\gamma te^{-\gamma t}\text{d}t = -\left.\frac{\Gamma(2,\gamma t)}{\gamma}\right\vert_{0}^{\infty} = \frac{1}{\gamma}
\end{equation}
The term $\Gamma(m,x)$ is the incomplete gamma function, and $l = \gamma$ so the normalized probability distribution is given as:
\begin{equation}
\label{Poisson3}
P(t) = \gamma^{2}te^{-\gamma t}
\end{equation}
Applying this equation with $\tau = 4\times 10^{6}$ yrs on average between impacts, the probability distribution function and cumulative distribution for the age of a randomly observed Pwyll-sized crater are computed and given in Figure \ref{fig:PwyllProb}.

From Figure \ref{fig:PwyllProb}, the probability that a given crater the size of Pwyll formed less than a given age is 2.7\% for 1 Myr, 50.0\% for 6.72 Myr, and 71.3\% for 10 Myr. There is negligible probability that the crater is older than 30 Myr or less than 0.1 Myr, and these results suggest with $\sim 97 \%$ probability that $t_{P} > 10^{6}$ years. Thus, it is possible but unlikely that Pwyll is young enough to explain the absence of a NSR signature without implying very slow NSR rates (see Figures \ref{fig:nstar_backwards_nonsat} and \ref{fig:nstar_backwards_sat}). These statistical results encourage a revisiting of the prior results using the observational models. Specifically, if we trust this simple statistical analysis, the lowest four curves corresponding to $t_{P} < 10^{6}$ yrs in Figures \ref{fig:nstar_backwards_nonsat} and \ref{fig:nstar_backwards_sat} can be excluded with 97\% confidence. The observational models thus predict a cosine signature in the albedo ratio data so large that it approaches the same order of magnitude as the sinusoidal signature for $n^{*} > 10^{-5}$ yr\textsuperscript{$-1$}. Note that as $|a_{1}/b_{1}| \rightarrow 1$, the deriving assumptions that the NSR signature is small break down, so a new analysis would be needed to determine the true scale of $a_{1}$ in such a case. However, that is unnecessary for this work, as the derivation assumptions fit the observed data quite well. Lastly, as long as $t_{P} \geq 10^{6}$ yrs, both observational models suggest a detectable NSR signal for $n^{*} > 10^{-7}$ yr\textsuperscript{$-1$} (conservative estimates of the limits of detectability are discussed in subsection 4.3). Since no detectable signal is found, we can conclude from the observational models that if $t_{P} > 10^{6}$ yrs, the NSR period is at least tens of millions of years.

\subsection{Error Analysis}
Because the analysis relies on a Fourier fit of the albedo ratio, a short statistical analysis of the fit error is useful for determining the resolution to which the Fourier coefficients are trustworthy. No error measures were provided with the original albedo ratio curve in \cite{Nelson_Icarus1986}. Let the residual error curve $e_{N}(\lambda)$ be computed by subtracting an $N$\textsuperscript{th} order Fourier fit of the original data from the curve of the original data. Comparing the original data to the Fourier fit for $N = 1$ using Figure \ref{fig:FFit}, we compute $e_{1}(\lambda)$, which is zero mean with a standard deviation of $\sigma_{e_{1}} = 0.024$, but exhibits some residual higher-order signal. Deterministic signals in $e_{N}(\lambda)$ persist as $N$ is increased until at least $N = 24$, corresponding to a fit of the data with a longitudinal angular resolution of $7.5^{\circ}$, and we compute $\sigma_{e_{24}}=0.014$. Treating this entire residual signal as noise should conservatively bound the effects of whatever the true noise is. 

The albedo ratio data $r(\lambda)$ may be written in the discrete form $r (\lambda_{j}) = f(\lambda_{j}) + n_{j}$, where $j = 0, 1, \ldots, D-1$, $ f(\lambda_{j})$ is the $N$\textsuperscript{th} order truncated Fourier fit in our case, and $e_{N}(\lambda)$ is discretized as $n_{j}$. Let the noise $n$ be random with mean $\overline{n}= 0$ and assumed variance $\sigma_{e_{N}}^{2}$. We may write the estimated Fourier coefficients in the following form demonstrated for the cosine coefficient, which is of the greatest concern: 
\begin{equation}
\label{error_ak1}
\hat{a}_{k} = \frac{2}{D}\sum_{j=0}^{D-1}f(\lambda_{j})\cos{k\lambda_{j}} + \frac{2}{D}\sum_{j=0}^{D-1}n_{j}\cos{k\lambda_{j}} = a_{k} + \varepsilon_{k}
\end{equation}
where $k$ corresponds to the coefficient order and $\lambda_{j} \in [0, 2\pi]$. The error in the Fourier coefficient is $\Delta a_{k} = \hat{a}_{k} - a_{k} = \varepsilon_{k}$ and we are interested in how the signal error statistics influence the Fourier coefficient error statistics: 
\begin{equation}
\label{error_ak2}
 \text{Var}(\varepsilon_{k}) = \sum_{j=0}^{D-1}\text{Var}\left(\frac{2n_{j}}{D}\cos{k\lambda_{j}}\right) = \frac{4\sigma_{e_{N}}^{2}}{D^{2}}\sum_{j=0}^{D-1}\cos^{2}{k\lambda_{j}} = \begin{cases}
      \frac{2\sigma_{e_{N}}^{2}}{D}, & k\neq0 \\
      \frac{4\sigma_{e_{N}}^{2}}{D}, & k=0
    \end{cases}
\end{equation}
Applying this result, we can bound the conservative error estimate in the Fourier coefficient with a high degree of certainty with the $3\sigma$ boundary:
\begin{equation}
\label{a_error}
|\Delta a_{1}| < 3\sqrt{\frac{2\sigma_{e_{N}}^{2}}{D}}
\end{equation}
Using this equation with $\sigma_{e_{24}} = 0.014$, we find $|\Delta a_{1}| < 2\times 10^{-3}$, thus the analysis suggests that the resolution in $a_{1}$ might be poor at and especially below the order of $10^{-3}$. Because the error analysis is conservative, so is this error result. The results presented earlier generally predict $a_{1} \geq \mathcal{O}(10^{-2})$ for $n^{*} \geq \mathcal{O}(10^{-5}) \ \text{yr}^{-1}$ except for an extremely young age of Pwyll crater (thousands of years or less) or a very thin homogeneous surface ice layer (1 mm or less), so the analyses using the models in this paper definitely yield the constraint $n^{*} < \mathcal{O}(10^{-5})$, even with this worst-case error analysis. The worst-case error only softens the constraint  $n^{*} < \mathcal{O}(10^{-6})$ in the case of an unlikely age of Pwyll and unlikely thickness of the homogeneous surface ice layer.

\subsection{Other Considerations}
Given the unexpected implication of our analysis that Europa is not in a state of NSR on the timescale predicted by other studies, it is useful to consider other possibilities and discuss their likelihoods. 
First, it is theoretically possible that some unmodeled process is removing the expected considerable cosine signature in the albedo ratio data predicted by the observational and physical models. In order to remove the NSR signature, some process with a cosine variation in longitude would be needed. There is no such known process given the general environmental symmetry about the line passing between $\lambda = 90^{\circ}$ E and $\lambda = 270^{\circ}$ E. Furthermore, if such a process did exist, there is no reason that it should be of exactly such a magnitude that it erases the NSR signature without leaving some residual cosine signature of its own. Therefore, in order to produce the observed albedo variations, an additional unknown process would need to oppose the well-understood radiolytic discoloration mechanism by a substantial margin. 

Additionally, it may be possible some sort of photometric or data-processing effect removed the cosine signature in the albedo data, but no such effects are mentioned in \cite{Nelson_Icarus1986}, and indeed the authors commented on the highly sinusoidal nature of the UV/V albedo ratio.

Another possibility is that Pwyll is unexpectedly young. In particular, if Pwyll is on the order of $10^{3}$ years old or younger, the analysis in Section 3.2 permits the resultant constant of the cosine signature $a_{1}$ to approach a small enough scale that it is hard to discern the resulting NSR signal from noise in the albedo ratio data from Voyager. However, it would still not explain the conclusions of the physical model (Section 3.3), which are similar to those of the observational models without relying on the age of Pwyll crater. 

If Pwyll is not unexpectedly young, perhaps it is possible that the icy ejecta from the Pwyll impact crater is in some way physically distinct from the surrounding ice, in such a way that the timescale of discoloration is slower for the crater ice than for surrounding ice. An investigation of this is beyond the scope of this work, and this possibility still would not explain the results of the physical model. 

While alternate explanations for the small ratio $a_{1}/b_{1}$ are possible, they are unlikely. Multiple independent approximate models of the surface discoloration process yield similar predictions for the scale of the cosine signature. For NSR with a period on the scale of $T^{*} < 10^{6}$, the crater Pwyll must be extremely young (tens of thousands of years old or less), or there must be an unknown systematic error in the Voyager albedo ratio data (or both).

%\newpage
\section{Conclusions}

The pervasive tectonic features on the surface of Europa have long been interpreted as evidence for accumulated stresses due to non-synchronous rotation (NSR) of the ice shell.  Previous studies have suggested a wide range of possible NSR rates for Europa. \cite{GreenbergEuropaRotation2002} review many past studies, highlighting the possibility of rates ranging from $n^{*}\sim10^{-8}$ yr\textsuperscript{$-1$} to $n^{*}\sim 10^{-5}$ or $n^{*}\sim 10^{-4}$ yr\textsuperscript{$-1$}, concluding that an NSR period of $50\, 000$ yrs to $250\, 000$ is likely, with an uncertainty factor of 5. Various other works agree with the conclusion that $n^{*}\sim 10^{-5}$ yr\textsuperscript{$-1$} \citep{Hoppa:Icarus2001, greenberg1998tectonic}. This work provides new constraints on the rate of NSR using the hemispheric color asymmetry, particularly the exogenic sinusoidal longitudinal variation in the ultraviolet to violet albedo ratio \citep{Nelson_Icarus1986}. No detectable NSR signature was found by this approach.

The analysis in this study relies on three models of surface discoloration. Two are lower fidelity ``observational" models of surface discoloration, and one is based on an approximate physical model of the dominant relevant processes acting on the surface ice. The results of the analysis using all three models reject an NSR rate on the order of $10^{-5}$ yr\textsuperscript{$-1$} as being highly unlikely, unless the crater Pwyll is an extremely recent surface feature, forming within only the last $\sim 10^{3}$ years, and the gardening-processed homogeneous contaminated surface ice layer is less than 1 mm thick, as opposed to the expected cm scale. The analysis also suggests that an NSR rate on the scale of $10^{-6}$ yr\textsuperscript{$-1$} is unlikely, unless Pwyll formed within the last $\sim 10^{4}$ years, and the homogeneous surface ice layer is less than 1 cm. Based on our error analysis, constraints to the NSR rate beyond this order of magnitude are potentially unreliable for the physical model. The observational model results still predict a discernible cosine signature for NSR periods up to $10^{7}$ years if Pwyll crater is more than $\sim 300 \, 000$ years old. We conclude using the available data of the hemispheric color dichotomy that Europa's NSR period, if it exists, is likely to be well over $10^{6}$ years -- a conclusion agreed upon by all three models.

How the aforementioned tectonic features could be explained given a multi-million year or slower NSR period is still unexplained. However, many tectonic features could be potentially explained by polar wander if it were a large reorientation about the tidal axis \citep{Greenberg2003TidalStress}. A sufficiently large reorientation in this direction could also effectively reset the longitudinal discoloration pattern near the equator. For example, a reorientation of $90^{\circ}$ would result in the new equator being composed of the great circle that once separated the leading and trailing hemispheres. The models in this paper predict that angular variations in albedo ratio along this great circle would be minimal, and the discoloration pattern would effectively be reset by such an event if it were sufficiently rapid. Polar reorientation by up to $90^{\circ}$ are predicted for latitudinally varying thickness of an ice shell on Europa \citep{Ojakangas1989EuropaPW, Ojakangas1989EuropaThermal,Matsuyama:2014,SchenkEA:Nature2008}. Our analysis provides no constraints on large reorientations mainly along the tidal axis, which have been predicted by some authors. For example, \cite{SchenkEA:Nature2008} suggested a $\sim 80^{\circ}$ true polar wander episode, primarily around the tidal axis, but with the potential for some longitudinal displacement as well. Their work suggested a small eastward shift in addition to the large tidal axis reorientation, but suggested other solutions were possible. Although our model cannot directly test for the history of large tidal axis reorientations, geologically recent small polar wander reorientations on the order of a few degrees, with an accompanying longitudinal shift, would be expected to leave a large and long-lasting cosine signature in the albedo ratio pattern, which is not observed.

In the future, the approach in this paper could be expanded to test for polar wander by looking for latitudinal signatures in the global ultraviolet/violet albedo ratio. That is not possible directly using the Voyager data from Nelson et al. (1986) due to limitations from the original data processing, and the fact that most of the albedo ratio data excludes the high latitudes. Additionally, UV spectra and visible or near-IR image data from Galileo could be combined to further investigate the hemispheric color dichotomy and its potential exogenous origins. Measurements at relevant wavelengths are also expected from Europa Clipper's ultraviolet spectrograph (UVS) \citep{retherford2018nasa} and imaging systems (EIS) \citep{turtle2019europa}, covering a range of longitudes \citep{pappalardo2019europa}. Thermal inertia variations might also be detected by the Europa Thermal Emission Imaging System (E-THEMIS) \citep{christensen2017looking} due to sintering of Europa's icy surface by high-energy particles, similar to those observed by Cassini on several icy Saturnian satellites \citep{paranicas2014lens}. Resolving such patterns and spectral behavior over a range of UV to thermal-IR wavelengths could provide further insights into the dichotomy's specific chemical nature and better constrain possible sources and discoloration rates. Additionally, approaches similar to the one introduced in this paper could be applied to the study of the numerous saturnian and jovian moons which exhibit signs of exogenic discoloration.

\subsection*{Data Availability}
\noindent The data that support the findings of this study are available from the authors on request.

\subsection*{Code Availability}
\noindent Any codes used for analysis in this paper are available from the authors on reasonable request.

\bibliographystyle{apalike}
\bibliography{references_PS.bib}   % Need to move the references below to a references file

\begin{thebibliography}{}

\bibitem[Bierhaus et~al., 2001]{BierhausIcarus2001}
Bierhaus, E.~B., Chapman, C.~R., Merline, W.~J., Brooks, S.~M., and Asphaug, E.
  (2001).
\newblock {Pwyll Secondaries and Other Small Craters on Europa}.
\newblock {\em Icarus}, 153:264--276.

\bibitem[Carlson et~al., 2009]{carlson2009europa}
Carlson, R., Calvin, W., Dalton, J., Hansen, G., Hudson, R., Johnson, R.,
  McCord, T., and Moore, M. (2009).
\newblock Europa's surface composition.
\newblock {\em Europa}, pages 283--327.

\bibitem[Christensen et~al., 2017]{christensen2017looking}
Christensen, P.~R., Hamilton, V.~E., Edwards, C.~S., and Spencer, J.~R. (2017).
\newblock {Looking Forward-A Next Generation of Thermal Infrared Planetary
  Instruments}.
\newblock In {\em AGU Fall Meeting Abstracts}, volume 2017, pages P33H--01.

\bibitem[Costello et~al., 2018]{COSTELLO2018327}
Costello, E.~S., Ghent, R.~R., and Lucey, P.~G. (2018).
\newblock The mixing of lunar regolith: Vital updates to a canonical model.
\newblock {\em Icarus}, 314:327 -- 344.

\bibitem[Costello et~al., 2020]{Costello:LPSC2020}
Costello, E.~S., Phillips, C.~B., Ghent, R.~R., and Lucey, P.~G. (2020).
\newblock {Impact Impacts on Europa's Uppermost Meters}.
\newblock In {\em 51st Lunar and Planetary Science Conference}, number 1565.

\bibitem[Geissler et~al., 1998]{Geissler:1998Nature}
Geissler, P.~E., Greenberg, R., Hoppa, G., Helfenstein, P., McEwen, A.,
  Pappalardo, R., Tufts, R., Ockert-Bell, M., Sullivan, R., Greeley, R.,
  Belton, M. J.~S., Denk, T., Clark, B., Burns, J., Veverka, J., and the
  Galileo Imaging~Team (1998).
\newblock {Evidence for non-synchronous rotation of Europa}.
\newblock {\em Nature}, 391(6665):368--370.

\bibitem[Goldreich and Peale, 1966]{GoldreichPeale:1966Nature}
Goldreich, P. and Peale, S. (1966).
\newblock {Resonant Spin States in the Solar System}.
\newblock {\em Nature}, 209:1078--1079.

\bibitem[Greeley et~al., 1998]{GREELEY1998}
Greeley, R., Sullivan, R., Klemaszewski, J., Homan, K., Head, J.~W.,
  Pappalardo, R.~T., Veverka, J., Clark, B.~E., Johnson, T.~V., Klaasen, K.~P.,
  Belton, M., Moore, J., Asphaug, E., Carr, M.~H., Neukum, G., Denk, T.,
  Chapman, C.~R., Pilcher, C.~B., Geissler, P.~E., Greenberg, R., and Tufts, R.
  (1998).
\newblock {Europa: Initial Galileo Geological Observations}.
\newblock {\em Icarus}, 135(1):4--24.

\bibitem[Greenberg et~al., 1998]{greenberg1998tectonic}
Greenberg, R., Geissler, P., Hoppa, G., Tufts, B.~R., Durda, D.~D., Pappalardo,
  R., Head, J.~W., Greeley, R., Sullivan, R., and Carr, M.~H. (1998).
\newblock {Tectonic processes on Europa: Tidal stresses, mechanical response,
  and visible features}.
\newblock {\em Icarus}, 135(1):64--78.

\bibitem[Greenberg et~al., 2003]{Greenberg2003TidalStress}
Greenberg, R., Hoppa, G.~V., Bart, G., and Hurford, T. (2003).
\newblock {Tidal Stress Patterns on Europa's Crust}.
\newblock {\em Celestial Mechanics and Dynamical Astronomy}, 87:171--188.

\bibitem[Greenberg et~al., 2002]{GreenbergEuropaRotation2002}
Greenberg, R., Hoppa, G.~V., Geissler, P., Sarid, A., and Tufts, B.~R. (2002).
\newblock {The Rotation of Europa}.
\newblock {\em Celestial Mechanics and Dynamical Astronomy}, 83:35--47.

\bibitem[Grundy et~al., 2007]{Grundy:2007Science}
Grundy, W.~M., Buratti, B.~J., Cheng, A.~F., Emery, J.~P., Lunsford, A.,
  McKinnon, W.~B., Moore, J.~M., Newman, S.~F., Olkin, C.~B., Reuter, D.~C.,
  Schenk, P.~M., Spencer, J.~R., Stern, S.~A., Throop, H.~B., and Weaver, H.~A.
  (2007).
\newblock {New Horizons Mapping of Europa and Ganymede}.
\newblock {\em Science}, 318(5848):234--237.

\bibitem[Hamilton, 1993]{HAMILTON1993244}
Hamilton, D.~P. (1993).
\newblock {Motion of Dust in a Planetary Magnetosphere: Orbit-Averaged
  Equations for Oblateness, Electromagnetic, and Radiation Forces with
  Application to Saturn's E Ring}.
\newblock {\em Icarus}, 101(2):244 -- 264.

\bibitem[Hoppa et~al., 2001]{Hoppa:Icarus2001}
Hoppa, G.~V., {Randall Tufts}, B., Greenberg, R., Hurford, T., O'Brien, D., and
  Geissler, P.~E. (2001).
\newblock {Europa's Rate of Rotation Derived from the Tectonic Sequence in the
  Astypalaea Region}.
\newblock {\em Icarus}, 153(1):208 -- 213.

\bibitem[Hoppa et~al., 1999]{HoppaScience1999}
Hoppa, G.~V., Tufts, B.~R., Greenberg, R., and Geissler, P.~E. (1999).
\newblock {Formation of Cycloidal Features on Europa}.
\newblock {\em Science}, 285:1899 -- 1902.

\bibitem[Howett et~al., 2012]{Howett:2012PacMan}
Howett, C. J.~A., Spencer, J.~R., Hurford, T., Verbiscer, A., and Segura, M.
  (2012).
\newblock {PacMan returns: An electron-generated thermal anomaly on Tethys}.
\newblock {\em Icarus}, 221:1084--1088.

\bibitem[{Johnson} et~al., 2004]{2004jpsm.book..485J}
{Johnson}, R.~E., {Carlson}, R.~W., {Cooper}, J.~F., {Paranicas}, C., {Moore},
  M.~H., and {Wong}, M.~C. (2004).
\newblock {\em {Radiation effects on the surfaces of the Galilean satellites}},
  volume~1, pages 485--512.

\bibitem[Johnson et~al., 1983]{Johnson_JGR1983}
Johnson, T.~V., Soderblum, L.~A., Mosher, J.~A., Danielson, G.~E., Cook, A.~F.,
  and Kupferman, P. (1983).
\newblock {Global Multispectral Mosaics of the Icy Galilean Satellites}.
\newblock {\em J. Geophysical Research}, 88(B7):5789 -- 5805.

\bibitem[Kattenhorn, 2002]{KATTENHORN2002490}
Kattenhorn, S.~A. (2002).
\newblock {Nonsynchronous Rotation Evidence and Fracture History in the Bright
  Plains Region, Europa}.
\newblock {\em Icarus}, 157(2):490 -- 506.

\bibitem[Levison et~al., 2000]{Levison:Icarus2000}
Levison, H.~F., Duncan, M.~J., Zahnle, K., Holman, M., and Dones, L. (2000).
\newblock {Planetary Impact Rates from Ecliptic Comets}.
\newblock {\em Icarus}, 143(2):415 -- 420.

\bibitem[Matsuyama et~al., 2014]{Matsuyama:2014}
Matsuyama, I., Nimmo, F., and Mitrovica, J.~X. (2014).
\newblock {Planetary Reorientation}.
\newblock {\em Annual Review of Earth and Planetary Sciences}, 42(1):605--634.

\bibitem[McEwen, 1986]{McEwen1986Europa}
McEwen, A.~S. (1986).
\newblock Exogenic and endogenic albedo and color patterns on europa.
\newblock {\em Journal of Geophysical Research}, 91(B8):8077--8097.

\bibitem[{McFadden} et~al., 1980]{McFadden1980}
{McFadden}, L., Bell, J., and {Mc Cord}, T. (1980).
\newblock {Visible} spectral reflectance measurements of the {Galilean}
  satellites at many orbital phase angles.
\newblock {\em Icarus}, 44(2):410--430.

\bibitem[Morrison and Burns, 1975]{MorrisonBurns1976}
Morrison, D. and Burns, J.~A. (1975).
\newblock The jovian satellites.
\newblock In {\em Jupiter: Studies of the interior, atmosphere, magnetosphere,
  and satellites}, pages 991--1034, Tucson, AZ. University of Arizona Press.

\bibitem[Nelson et~al., 1986]{Nelson_Icarus1986}
Nelson, M.~L., McCord, T.~B., Clark, R.~N., Johnson, T.~V., Matson, D.~L.,
  Mosher, J.~A., and Soderblom, L.~A. (1986).
\newblock {Europa: Characterization and Interpretation of Global Spectral
  Surface Units}.
\newblock {\em Icarus}, 65:129--151.

\bibitem[Nimmo, 2004]{Nimmo2004}
Nimmo, F. (2004).
\newblock Stresses generated in cooling viscoelastic ice shells: Application to
  europa.
\newblock {\em Journal of Geophysical Research: Planets}, 109(E12).

\bibitem[Nimmo et~al., 2007]{nimmo2007global}
Nimmo, F., Thomas, P., Pappalardo, R., and Moore, W. (2007).
\newblock {The global shape of Europa: Constraints on lateral shell thickness
  variations}.
\newblock {\em Icarus}, 191(1):183--192.

\bibitem[Ojakangas and Stevenson, 1989a]{Ojakangas1989EuropaPW}
Ojakangas, G.~W. and Stevenson, D.~J. (1989a).
\newblock {Polar wander of an Ice Shell on Europa}.
\newblock {\em Icarus}, 81:242--270.

\bibitem[Ojakangas and Stevenson, 1989b]{Ojakangas1989EuropaThermal}
Ojakangas, G.~W. and Stevenson, D.~J. (1989b).
\newblock {Thermal State of an Ice Shell on Europa}.
\newblock {\em Icarus}, 81:220--241.

\bibitem[Pappalardo et~al., 2019]{pappalardo2019europa}
Pappalardo, R.~T., Senske, D., Korth, H., Becker, T.~M., Blaney, D.~L.,
  Blankenship, D.~D., Christensen, P.~R., Gudipati, M.~S., Hayes, A., Kempf,
  S., et~al. (2019).
\newblock {The Europa Clipper: Science and Mission}.
\newblock In {\em AGU Fall Meeting Abstracts}, volume 2019, pages P53B--06.

\bibitem[{Paranicas} et~al., 2009]{2009euro.book..529P}
{Paranicas}, C., {Cooper}, J.~F., {Garrett}, H.~B., {Johnson}, R.~E., and
  {Sturner}, S.~J. (2009).
\newblock {\em {Europa's Radiation Environment and Its Effects on the
  Surface}}, page 529.

\bibitem[Paranicas et~al., 2014]{paranicas2014lens}
Paranicas, C., Roussos, E., Decker, R., Johnson, R., Hendrix, A., Schenk, P.,
  Cassidy, T., Dalton~III, J., Howett, C., Kollmann, P., et~al. (2014).
\newblock {The lens feature on the inner saturnian satellites}.
\newblock {\em Icarus}, 234:155--161.

\bibitem[Retherford et~al., 2018]{retherford2018nasa}
Retherford, K., Feldman, P.~D., Spencer, J., Davis, M., Gladstone, R., Saur,
  J., Stern, A., Raut, U., Greathouse, T., Versteeg, M., et~al. (2018).
\newblock {NASA's Europa Clipper Mission Ultraviolet Spectrograph
  (Europa-UVS)}.
\newblock {\em 42nd COSPAR Scientific Assembly}, 42:B5--3.

\bibitem[Sarid et~al., 2002]{Sarid2002PolarWander}
Sarid, A.~R., Greenberg, R., Hoppa, G.~V., Hurford, T.~A., Tufts, B.~R., and
  Geissler, P. (2002).
\newblock {Polar wander and surface convergence on Europa: Evidence from a
  survey of strike-slip displacement}.
\newblock {\em Icarus}, 158:24--41.

\bibitem[Schenk et~al., 2011]{Schenk:2011PlasmaSaturn}
Schenk, P., Hamilton, D.~P., Johnson, R.~E., McKinnon, W.~B., Paranicas, C.,
  Schmidt, J., and Showalter, M.~R. (2011).
\newblock Plasma, plumes and rings: Saturn system dynamics as recorded in
  global color patterns on its midsize icy satellites.
\newblock {\em Icarus}, 211:740--757.

\bibitem[Schenk et~al., 2008]{SchenkEA:Nature2008}
Schenk, P., Matsuyama, I., and Nimmo, F. (2008).
\newblock {True polar wander on Europa from global-scale small-circle
  depressions}.
\newblock {\em Nature}, 453(7193):368--371.

\bibitem[Trumbo et~al., 2019]{Trumbo_2019}
Trumbo, S.~K., Brown, M.~E., and Hand, K.~P. (2019).
\newblock {H2O2 within Chaos Terrain on Europa's Leading Hemisphere}.
\newblock {\em The Astronomical Journal}, 158(3):127.

\bibitem[Turtle et~al., 2019]{turtle2019europa}
Turtle, E., McEwen, A., Bland, M., Collins, G., Daubar, I., Ernst, C.,
  Fletcher, L., Hansen, C., Hawkins, A.~H., Humm, D., et~al. (2019).
\newblock {The Europa Imaging System (EIS): High-resolution, 3-D insight into
  Europa's geology, ice shell, and potential for current activity}.
\newblock {\em RED}, 520:590.

\end{thebibliography}

\end{document}